%% file: main.tex
\pdfoutput=1

\documentclass[12pt]{article}

\newcommand{\blind}{0}

\usepackage{amsmath}
\usepackage{amssymb}
\usepackage{amsthm}
\usepackage{bm}
\usepackage{mathtools}
\usepackage{caption}
\usepackage[a4paper, margin=1in]{geometry}
\usepackage{graphicx}
\usepackage[colorlinks, linkcolor=blue, citecolor=blue, urlcolor=blue, bookmarks=false, hyperfootnotes=false]{hyperref}
\usepackage{multibib}
\usepackage{natbib}
\usepackage{setspace}
\usepackage{titling}

\usepackage{subfiles}

\theoremstyle{plain}
\newtheorem{theorem}{Theorem}[section]
\newtheorem{corollary}[theorem]{Corollary}

\theoremstyle{definition}

\newtheorem{remark}[theorem]{Remark}

\newcites{pref}{References}
\newcites{sref}{Supplemental References}

\newcommand{\papertitle}{\Large \bf Survey Design and Estimating Equations when Combining Big Data with Probability Samples}
\newcommand{\paperauthors}{Ryan Covey,\protect\footnotemark[1] \protect\footnotemark[4]\, Lucca Buonamano\protect\footnotemark[2] \protect\footnotemark[4]}
\newcommand{\mdaffiliation}{Methodology and Data Science Division, Australian Bureau of Statistics}
\newcommand{\mddisclaimer}{We would like to thank Siu-Ming Tam, Lyndon Ang, Anders Holmberg, Paul Schubert, Bernadette Giuffrida, Tim Cadogan-Cowper, Qinghuan Luo and Diego Zardetto for useful comments and suggestions leading to an improved manuscript. Views expressed in this paper are those of the authors and do not necessarily represent those of the Australian Bureau of Statistics. Where quoted or used, they should be attributed clearly to the authors.}
\newcommand{\ryanfootnote}{\mdaffiliation, Level 3, 818 Bourke Street, Docklands, VIC 3000, Australia. Email: ryan.covey@abs.gov.au}
\newcommand{\luccafootnote}{\mdaffiliation, Level 8, 44 Market Street, Sydney, NSW 2000, Australia. Email: lucca.buonamano@abs.gov.au}
\newcommand{\mydate}{August 9, 2023}

\bibliographystylepref{chicago}

\newcommand{\wholepaperandsupplemental}{0}
\newcommand{\wholepaper}{0}

\begin{document}
\renewcommand{\wholepaperandsupplemental}{1}

\subfile{title_page}

\subfile{paper}

\subfile{supplemental_data_title_page}

\subfile{supplemental}

\end{document}

%% file: title_page.tex
\begin{titlingpage}

\if0\blind
{
	\renewcommand{\thefootnote}{\fnsymbol{footnote}}
	\title{\papertitle}
	\author{\paperauthors}
	\date{\mydate}
	\maketitle
	\footnotetext[1]{\ryanfootnote}
	\footnotetext[2]{\luccafootnote}
	\footnotetext[4]{\mddisclaimer}
	\setcounter{footnote}{0}
	\renewcommand{\thefootnote}{\arabic{footnote}}
} \fi

\begin{abstract}
The use of big data in official statistics and the applied sciences is accelerating, but statistics computed using only big data often suffer from substantial selection bias. This leads to inaccurate estimation and invalid statistical inference. We rectify the issue for a broad class of linear and nonlinear statistics by producing estimating equations that combine big data with a probability sample. Under weak assumptions about an unknown superpopulation, we show that our integrated estimator is consistent and asymptotically unbiased with an asymptotic normal distribution. Variance estimators with respect to both the sampling design alone and jointly with the superpopulation are obtained at once using a single, unified theoretical approach. A surprising corollary is that strategies minimising the design variance almost minimise the joint variance when the population and sample sizes are large. The integrated estimator is shown to be more efficient than its survey-only counterpart if dependence between sample membership indicators is small and the finite population is large. We illustrate our method for quantiles, the Gini index, linear regression coefficients and maximum likelihood estimators where the sampling design is stratified simple random sampling without replacement. Our results are illustrated in a simulation of individual Australian incomes.
\end{abstract}


\end{titlingpage}

%% file: paper.tex
\renewcommand{\wholepaper}{1}

\doublespacing

\subfile{introduction/introduction.tex}

\subfile{integrated_estimation/integrated_estimation.tex}

\subfile{theoretical_results/theoretical_results.tex}



\subfile{examples/examples.tex}

\subfile{simulation/simulation.tex}


\subfile{conclusion/conclusion.tex}

\bibliographypref{library}

\newpage

%% file: introduction/introduction.tex
\section{Introduction}

Big data is taking on an increasingly dominant role in both official statistics and empirical research, but big data also introduces many problems for estimation and inference that are not present when relying only on probability surveys or censuses \citeppref{Christen2022}. Despite concerns about the representativeness, sustainability, relevance and interpretability of big data, a number of factors are nevertheless increasing the adoption of big data by National Statistics Offices (NSOs) \citeppref{Holmberg2012, Citro2014, Tam2015, Meng2018}. These factors include declining response rates, the high respondent burden of surveys, the affordability (to the NSO) of big data collection relative to survey collection and the demand for more timely and comprehensive statistical releases servicing both traditional and emerging data needs. There are also many practical challenges for NSOs to overcome in order to effectively link big and survey data so that their information may be combined; see \citepref{Lothian2019}.

In this paper, we consider estimation and inference using \textit{estimating equations} that integrate big and survey data to improve accuracy but preserve asymptotic unbiasedness. We cater to a broad class of statistics that includes the mean, the median and other quantiles, the Gini coefficient, linear regression coefficients, maximum likelihood estimators, and many others; see Chapter 5 of \citepref{vanderVaart1998} for a text-book introduction. We extend \citepref{Binder1983} and \citepref{Godambe1986} to provide estimators of the variance under the survey design alone (\textit{design variance}, for finite population inference) and jointly with the unknown superpopulation (\textit{joint variance}, for inference about superpopulation parameters) produced in part by standard estimators of the design variance for sums and averages (e.g.\ \citealppref{Sarndal1992}). Our approach is therefore applicable to the same complex survey designs for which standard design-based variance estimators are available. The joint variance is equal to the anticipated variance of \citepref{Isaki1982} when the latter is taken with respect to the true superpopulation, and we accommodate superpopulation models that are incorrectly specified. We also incorporate the weights into our joint variance estimator to allow for informative sampling with nonignorable designs; see \citepref{Pfeffermann1993}.

Extending the approach of \citepref{Kim2021}, our integrated estimator is asymptotically unbiased with a smaller variance than the corresponding estimator produced without big data. Following \citepref{Lohr2021}, we treat the big data as a completely enumerated stratum to show how our design-based variance estimators can be used for optimal survey design in the presence of big data to target nonlinear parameters, leading to more accurate statistics with a lower cost. Further, we accommodate modifications to the standard Horvitz-Thompson weights, for example to account for nonresponse (e.g.\ \citealppref{Brick2013}) and of course to incorporate the big data.

A key advantage of our method is its general applicability. Existing approaches tend to apply to a narrower class of statistics; address variance from the design or superpopulation, but not both; cater only to particular sample designs; assume Horvitz-Thompson weights; or provide only informal justification: see references above, \citepref{Binder1995}, \citepref{Imbens1996}, \citepref{Kovacevic1997}, \citepref{Wooldridge1999, Wooldridge2001, Wooldridge2002}, \citepref{Bhattacharya2007} and \citepref{Lumley2017}. Aside from the integrated median of \citepref{Covey2023}, it appears to us that the literature on integrating big and survey data has so far ignored nonlinear statistics; see \citepref{Rao2021} and \citepref{Wu2022} for reviews.

In Section \ref{sec:intest}, we introduce our integrated estimator alongside its unintegrated counterpart, and outline some desirable properties that define the broader class of estimators we consider. Theoretical results are present in Section \ref{sec:theory}, where we: 1) show that these estimators are close to their population and superpopulation counterparts for large population sizes, 2) provide estimators for both the design and joint variance based on large-population central limit theorems, and 3) show that normalising the weights so that they sum to one often (but not always) has no effect on the asymptotic behaviour of the estimator. These results are applied to produce variance estimators for examples in Section \ref{sec:eg}, where we consider Horvitz-Thompson and integrated weights, sample designs that use stratified simple random sampling without replacement, quantiles, the Gini index, linear regression coefficients and maximum likelihood estimators. In Section \ref{sec:sim}, we compare the performance of integrated, survey-only and big-data-only estimates of the median and Gini index in a simulation of Australian incomes. We finish with some concluding remarks in Section \ref{sec:conclusion}.

We will observe the following notational conventions. The indicator function $I(E)$ evaluates to one or zero according to whether or not the event $E$ is true. All probability statements, expectations and variances are under the joint distribution spanning both the survey design and superpopulation. This means that standard results in asymptotic statistics can be applied immediately, as written and without translation; see for example \citepref{vanderVaart1998} and \citepref{Davidson2021}. Probabilities, expectations and variances with respect to the design alone are then obtained via conditional probability; see Section \ref{subsec:theoryoverview}. We use $X_N \overset{P}{\to} X$ to denote convergence in probability of $X_N$ to $X$ as $N \to \infty$, and use $X_N = o_P(r_N)$ and $X_N = O_P(r_N)$ to express that $X_N$ converges to zero in probability or is bounded in probability according to the ``rate'' $r_N$; see Chapter 2 of \citepref{vanderVaart1998}.

\if0\wholepaper {
	\bibliographypref{../library}
} \fi

%% file: integrated_estimation/integrated_estimation.tex
\section{Integrated Estimation} \label{sec:intest}

Suppose we are interested in a $d$-dimensional column vector of population parameters $\hat{\theta}_N$ that is a function of a finite population $Y_1, Y_2, \ldots, Y_N$ of $N$ $k$-dimensional column vectors. We will assume that there exists a vector-valued function $\psi$ such that the population estimating equation
\begin{equation*}
\Psi_N(\theta) \coloneqq \frac{1}{N} \sum_{i=1}^N \psi(Y_i ; \theta)
\end{equation*}
is close to zero if and only if $\theta$ is close to $\hat{\theta}_N$.\footnote{The precise mathematical meaning of ``close to'' depends on what we are trying to prove; see Theorems \ref{thm:cons} and \ref{thm:normsup}.} There is such an equation for many different population parameters, including the mean ($\psi(y ; \theta) = y - \theta$), the median ($\psi(y ; \theta) = I(y < \theta) - I(y > \theta)$; see Section \ref{eg:quant}), the Gini index, linear regression coefficients and those produced via maximum likelihood estimation (see Sections \ref{eg:gini}, \ref{eg:lr} and \ref{eg:mle}, respectively).

In official statistics obtaining the entire population is often infeasible. In this case, we must contend with a sample statistic $\hat{\theta}_s$, defined so that $\Psi_s(\theta) \approx 0$ if and only if the function argument satisfies $\theta \approx \hat{\theta}_s$, where
\begin{equation}
\Psi_s(\theta) \coloneqq \frac{1}{N} \sum_{i=1}^N w_i \psi(Y_i ; \theta) \label{eqn:weightedzfunction}
\end{equation}
is the sample estimating equation for a given sequence of weights $w_1, w_2, \ldots, w_N$. The sample $s = \{i : I(w_i \neq 0)\}$ contains indices identifying observed units; zero-weighted units do not appear in \eqref{eqn:weightedzfunction}, and we also assume that they do not influence $\hat{\theta}_s$ and therefore do not require observation. The sample mean and linear regression coefficients can be defined by first defining $\psi$, then setting the sample estimating equation in \eqref{eqn:weightedzfunction} to zero and solving for $\theta$. On the other hand, the sample median and Gini index are usually defined in some other way, and an appropriate $\psi$ must be found after. The latter case is not easily accommodated by standard references such as \citepref{Godambe1984}. See Section \ref{sec:eg} for details.


If a probability survey is conducted, let the sample membership indicator $\alpha_i$ equal one or zero according to whether or not unit $i$ is observed, and let $\pi_i = \mathrm{Pr}(\alpha_i = 1 \mid Y_{1:N})$ be the first-order inclusion probability, where $Y_{1:N} = (Y_1, Y_2, \ldots, Y_N)$ contains all study variables for the entire population. Then the Horvitz-Thompson weights $w^{HT}_i = \alpha_i \pi_i^{-1}$ are a standard choice; see Chapter 2.8 of \citepref{Sarndal1992}. Let $\Psi^{HT}_s$ be the sample estimating equation obtained after substituting $w_i = w^{HT}_i$ into \eqref{eqn:weightedzfunction}. Since $\mathbb{E}[w_i^{HT} \mid Y_{1:N}] = \mathbb{E}[\alpha_i \mid Y_{1:N}] \pi_i^{-1} = \pi_i \pi_i^{-1} = 1$, $\Psi^{HT}_s(\theta)$ is \textit{design unbiased}, meaning that
\begin{equation}
\mathbb{E}[\Psi^{HT}_s(\theta) \mid Y_{1:N}] = \frac{1}{N} \sum_{i=1}^N \mathbb{E}[w^{HT}_i \mid Y_{1:N}] \psi(Y_i ; \theta) = \Psi_N(\theta). \label{eqn:htunbiased}
\end{equation} We will see in subsequent sections that design-unbiasedness of the sample estimating equation helps ensure that the sample estimator $\hat{\theta}_s$ is close to its population counterpart, $\hat{\theta}_N$. We also rely on design-unbiasedness for some aspects of variance estimation.


Now suppose that, in addition to the probability survey, we have access to a big data set with an unknown selection mechanism that is produced and maintained outside of an NSO for non-statistical purposes. Let $\delta_i$ equal one or zero according to whether or not $Y_i$ is observed through the big data set. If the big data is used naively and we adopt $\delta_1, \delta_2, \ldots, \delta_N$ as our weights, then it is likely that $\mathbb{E}[\delta_i \mid Y_{1:N}] \neq 1$ and in view of \eqref{eqn:htunbiased} $\Psi_s(\theta)$ is not necessarily design unbiased. On the other hand, if we follow \citepref{Kim2021} to adopt the Data Integrated weights $w^{DI}_i = \delta_i + (1 - \delta_i) w^{HT}_i$, then
\begin{equation}
\mathbb{E}[w^{DI}_i \mid Y_{1:N}] = \delta_i + (1 - \delta_i) \mathbb{E}[w^{HT}_i \mid Y_{1:N}] = \delta_i + 1 - \delta_i = 1, \label{eqn:diunbiased}
\end{equation}
where we extend the vector $Y_i$ so that its final element is equal to $\delta_i$. This leads us to condition on $\delta_{1:N} = (\delta_1, \ldots, \delta_N)$ and justifies the first equality in \eqref{eqn:diunbiased} above. Letting $\Psi^{DI}_s(\theta)$ denote the sample estimating equation for weights $w^{DI}_i$, we have $\mathbb{E}[\Psi^{DI}_s(\theta) \mid Y_{1:N}] = \Psi_N(\theta)$, and design-unbiasedness is retained. Notice that $\mathbb{E}[w^{DI}_i \mid Y_{1:N}] = 1$ because $\mathbb{E}[w^{HT}_i \mid Y_{1:N}] = 1$, and therefore design unbiasedness does not rely on any other characteristic of the Horvitz-Thompson weights specifically. As a result, we will from now on make use of the following generalisation of the weights of \citepref{Kim2021}:
\begin{equation}
w^{DI}_i = \delta_i + (1 - \delta_i) w_i, \label{eqn:wdi}
\end{equation}
where $w_i$ generically represents any set of weights satisfying $\mathbb{E}[w_i \mid Y_{1:N}] = 1$ (note that the notation $w_i$ can therefore be used to represent $w^{DI}_i$ itself). In a slight abuse of terminology, we will say that weights with this property are design-unbiased too, since design-unbiased weights lead to a design-unbiased sample estimating equation.

The generalisation of the integrated weights given in \eqref{eqn:wdi}  makes the integrated estimator $\hat{\theta}^{DI}_s$ applicable to survey weights that are produced by making adjustments to the Horvitz-Thompson weights, for example to account for survey nonresponse (e.g.\ \citealppref{Brick2013}). In Section \ref{eg:diweights} we show that $\hat{\theta}^{DI}_s$ is asymptotically unbiased, and that if the dependence between weights $w_i$ (induced by without-replacement sampling, say) is small enough, the integrated estimator $\hat{\theta}^{DI}_s$ has a smaller variance than the estimator $\hat{\theta}_s$ produced without the big data.

\if0\wholepaper {
	\bibliographypref{../library}
} \fi

%% file: theoretical_results/theoretical_results.tex
\section{Theoretical Results} \label{sec:theory}

\subsection{Overview} \label{subsec:theoryoverview}

In the design-based survey framework the population is fixed but unknown and all randomness is attributed to the weights (e.g.\ Chapter 2 of \citealppref{Sarndal1992}). As a result, the design-based variance of $\hat{\theta}_s$ reflects uncertainty related to the survey design, and only the survey design. This is appropriate when choosing between competing designs, where the aim is usually to minimise the design-based variance subject to constraints (for example, to allocate sample to strata as in Chapter 3.7.3 of \citealppref{Sarndal1992}), and we will see in Section \ref{sec:varest} that for large population and sample sizes the sample design minimising the design variance almost minimises the joint variance. By treating the target variables as nonrandom, we also avoid making assumptions that are unnecessary for the purposes of survey design, which can therefore proceed without the need for special expertise regarding the nature and behaviour of the target variables.

When working with nonlinear statistics, however, the design-based framework alone is insufficient to develop the kinds of asymptotic results provided in Section \ref{sec:cons} that justify the use of $\hat{\theta}_s$ to approximate $\hat{\theta}_N$ (nonlinear statistics are often biased). On its own, the design-based framework is also insufficient to develop the kinds of asymptotic results provided in Sections \ref{sec:normality} and \ref{sec:varest} that justify the use of design-based variance estimators to compare survey designs that aim to produce an accurate $\hat{\theta}_s$ in a cost effective manner. Further, scientific inquiry often seeks a $p$ value measuring the strength of the evidence in survey results against a null hypothesis related to some aspect of a presumed superpopulation model that is postulated to generate the population (e.g.\ \citealppref{Wooldridge1999, Wooldridge2001, Wooldridge2002, Bhattacharya2007}). In this scenario, randomness from \textit{both} the design \textit{and} the superpopulation need to be quantified in order to produce the $p$ value and conduct the hypothesis test, and here too a purely design-based approach is insufficient.

To resolve this apparent contradiction and cater to both uses of variance estimation, we will assume that there is some unknown superpopulation model that generates the population, but in our asymptotic results we rely only on weak assumptions that place little restriction on the form that the superpopulation model might take. Design-based expectations and variances are recovered using the law of total expectation by conditioning on the population $Y_{1:N}$, as seen in Section \ref{sec:intest} above. We also show in Section \ref{sec:varest} that if the weights are design unbiased, then the design variance of $\hat{\theta}_s$ is well approximated by the joint variance of $\hat{\theta}_s - \hat{\theta}_N$, which follows from a central limit theorem (presented in Section \ref{sec:normality}) derived under standard assumptions due to standard results in large-sample asymptotics (though `large-population' might be a more appropriate term in our context). Additionally, we can obtain an estimate of the joint variance of $\hat{\theta}_s$ simply by adding a design-based variance estimate for $\hat{\theta}_s$ to an estimate of the variance of $\hat{\theta}_N$ under the superpopulation. These results apply as $N$ approaches infinity for a given, fixed sampling fraction, and in practice we use the approximation that applies to the sampling fraction of the design at hand.

When producing a weighted average, one needs to decide whether or not to normalise the weights so that their average is equal to one. In Section \ref{sec:normweights}, we show that if $\hat{\theta}_s$ is constructed using the weights $w$ and the function $\psi$ so that $\Psi_s(\hat{\theta}_s) \approx 0$ (as it is for linear regression and maximum likelihood estimation; see Section \ref{eg:lr} and \ref{eg:mle}, respectively), then the normalisation (or not) of the weights has no impact on the asymptotic behaviour of $\hat{\theta}_s$ subject to the sufficient conditions of the preceding subsections. Note that this is not necessarily true if $\hat{\theta}_s$ is constructed using the weights in some other way. Quantiles for example should be produced using normalised weights; see Section \ref{eg:quant}. On the other hand, even for estimators that should be produced with normalised weights, we can use either normalised or unnormalised weights to construct $\Psi_s$.

Proofs for all results are provided in the supplementary appendix.

\subsection{Consistency} \label{sec:cons}

The following theorem establishes conditions ensuring that the sample estimator $\hat{\theta}_s$ converges in probability to a fixed $\theta_0$ as $N \to \infty$ under the joint distribution spanning both the design and superpopulation. In scientific inquiry $\theta_0$ often represents a key characteristic of the superpopulation being studied. If $w_i = 1$ then $\hat{\theta}_N =\hat{\theta}_s$, so this theorem can also be applied to show that $\hat{\theta}_N$ converges to $\theta_0$.

\begin{theorem} \label{thm:cons}
Assume that $\hat{\theta}_s$ satisfies
\begin{equation}
\Psi_s(\hat{\theta}_s) \overset{P}{\to} 0. \label{eqn:conszest}
\end{equation}
Also suppose there exists a fixed parameter vector $\theta_0$ and a fixed vector-valued function $\Psi$ such that for all $\epsilon > 0$,
\begin{gather}
\sup_{\theta} \left\lVert \Psi_s(\theta) - \Psi(\theta) \right\rVert \overset{P}{\to} 0, \label{eqn:consulln} \\
\inf_{\theta : \lVert \theta - \theta_0 \rVert > \epsilon} \lVert \Psi(\theta) \rVert > 0 = \lVert \Psi(\theta_0) \rVert. \label{eqn:consident}
\end{gather}
Then $\hat{\theta}_s$ converges in probability to $\theta_0$.
\end{theorem}

\begin{remark} \label{rmk:changingweights}
Note that each weight $w_i$ is permitted to vary with $N$, even though this relationship is suppressed in the notation. This is necessary for Theorem \ref{thm:cons} to be applicable to simple random sampling without replacement, which is a popular sampling scheme covered in Section \ref{eg:srswor}. To see this, consider taking such a sample of size $n = 5$ from $Y_1, Y_2, \ldots, Y_{10}$ ($N = 10$) and of size $n = 10$ from $Y_1, Y_2, \ldots, Y_{20}$ ($N = 20$). Recall that a unit $i$ is selected if and only if $w_i \neq 0$. If $w_i$ were not permitted to vary with $N$, then the second sample would certainly contain the first sample. This is false if simple random sampling without replacement is used to draw the second sample.
\end{remark}

\begin{remark} \label{rmk:thetadef}
When establishing consistency, \eqref{eqn:conszest} is the formal counterpart to the informal statement $\Psi_s(\hat{\theta}_s) \approx 0$. This assumption must be explicitly established in order to correctly define $\psi$ and $\theta_0$, if desired. Defining $\psi$ takes on a greater role in variance estimation, which we discuss the next subsection.
\end{remark}

\begin{remark}
The uniform law of large numbers in \eqref{eqn:consulln} extends the standard, pointwise law of large numbers $\Psi_s(\theta) \overset{P}{\to} \Psi(\theta)$ to ensure that the error has an upper bound across $\theta$. If this holds, the set of random variables obtainable by evaluating $w_i \psi(Y_i ; \theta)$ at a given $\theta$ is said to be a \textit{Glivenko-Cantelli class}; see Chapter 2.1 of \citepref{vanderVaart1996}. Alternatively, Chapter 22.5 of \citepref{Davidson2021} covers uniform laws of large numbers in a way that better accommodates dependence between both weights $w_i$ and observations $Y_i$. If $Y_i$ is independent and identically distributed (i.i.d.), then for any reasonable choice of weights, $\Psi_s(\theta) \overset{P}{\to} \mathbb{E}[\psi(Y_i ; \theta)]$ if the expectation exists. Therefore $\Psi(\theta) = \mathbb{E}[\psi(Y_i ; \theta)]$, which we can often use to obtain a workable expression for $\Psi(\theta)$ that can be used for variance estimation. This is illustrated in Section \ref{eg:stats}.
\end{remark}

\begin{remark}
The assumption given in $\eqref{eqn:consident}$ ensures that there is only one $\theta_0$ satisfying $\Psi(\theta_0) = 0$ to which $\hat{\theta}_s$ might converge, and that $\Psi(\theta)$ being close to zero implies that $\theta$ is close to $\theta_0$. Without this requirement, a $\hat{\theta}_s$ satisfying $\Psi_s(\hat{\theta}_s) \approx 0$ might be close to any number of dispersed $\theta$ satisfying $\Psi(\theta) \approx 0$, making it difficult to show convergence. In econometrics this requirement is often called \textit{identification}; see \citepref{Lewbel2019} for a review.
\end{remark}

In the corollary below, we show that only mild assumptions are required for the population estimator $\hat{\theta}_N$ to be well-approximated by the sample estimator $\hat{\theta}_s$. This interpretation is better-suited to survey design problems than it is to scientific inquiry, and doesn't require any knowledge of the form that $\theta_0$ or the superpopulation might take. 

\begin{corollary} \label{crl:conspop}
Suppose that the assumptions of Theorem \ref{thm:cons} are satisfied for the weights $w_1, w_2, \ldots, w_N$ and $1, 1, \ldots, 1$. If
\begin{equation}
\frac{1}{N} \sum_{i=1}^N (w_i - 1) \psi(Y_i ; \theta) \overset{P}{\to} 0 \label{eqn:llnwm1}
\end{equation}
for all $\theta$, then $\Psi_s(\theta)$ and $\Psi_N(\theta)$ both converge to the same $\Psi(\theta)$, and $\hat{\theta}_s$ and $\hat{\theta}_N$ both converge to the same $\theta_0$.
\end{corollary}

\begin{remark}
If the weights $w_1, w_2, \ldots, w_N$ are design-unbiased and $\mathbb{E}[\psi(Y_i ; \theta)]$ exists, then $\mathbb{E}[\frac{1}{N} \sum_{i=1}^N (w_i - 1) \psi(Y_i ; \theta)] = 0$, and a law of large numbers alongside asymptotic uniform integrability of the average (a technical condition; see Section 2.5 of \citealppref{vanderVaart1998}) gives \eqref{eqn:llnwm1}.
\end{remark}

\subsection{Asymptotic Normality} \label{sec:normality}

The following is a central limit theorem for $\sqrt{N}(\hat{\theta}_s - \theta_0)$ to converge in distribution, usually to a zero-mean normal distribution, as $N \to \infty$ under the joint distribution spanning both the design and superpopulation. It is most useful when conducting hypothesis tests and producing confidence intervals as part of a scientific inquiry. Recall that if $w_i = 1$ then $\hat{\theta}_N = \hat{\theta}_s$, so this theorem can also be applied to obtain the asymptotic distribution of $\hat{\theta}_N$.

\begin{theorem} \label{thm:normsup}
Assume that $\hat{\theta}_s \overset{P}{\to} \theta_0$ for some fixed $\theta_0$, and $\Psi_s(\hat{\theta}_s) = o_P(N^{-1/2})$. Suppose there exists a fixed vector-valued function $\Psi$ such that $\Psi(\theta_0) = 0$, with a continuously invertible Jacobian matrix $\dot{\Psi}_{\theta_0}$ at $\theta_0$, and
\begin{equation}
\sqrt{N}(\Psi_s(\hat{\theta}_s) - \Psi(\hat{\theta}_s)) - \sqrt{N}(\Psi_s(\theta_0) - \Psi(\theta_0)) = o_P(1 + \sqrt{N} \lVert \hat{\theta}_s - \theta_0 \rVert ). \label{eqn:donsker}
\end{equation}
Then we have
\begin{equation}
\sqrt{N} \dot{\Psi}_{\theta_0} (\hat{\theta}_s - \theta_0) = -\sqrt{N}(\Psi_s(\theta_0) - \Psi(\theta_0)) + o_P(1). \label{eqn:linsup}
\end{equation}
If we also have that $\sqrt{N}(\Psi_s(\theta_0) - \Psi(\theta_0)) \Rightarrow Z_s$ for some fixed random variable $Z_s$, then
\begin{equation}
\sqrt{N}(\hat{\theta}_s - \theta_0) \Rightarrow \dot{\Psi}_{\theta_0}^{-1} Z_s. \label{eqn:zestlin}
\end{equation}
\end{theorem}

\begin{remark}
In this theorem, $\Psi_s(\hat{\theta}_s) = o_P(N^{-1/2})$ is the formal counterpart to the informal statement $\Psi_s(\hat{\theta}_s) \approx 0$. This is a stronger requirement than \eqref{eqn:conszest}, which is assumed in Theorem \ref{thm:cons} to establish consistency; see Remark \ref{rmk:thetadef}. It is important to explicitly establish this assumption in order to correctly define $\psi$, which is needed for variance estimation later in this section and in the examples of Section \ref{sec:eg}.
\end{remark}

\begin{remark}
Note again that each weight $w_i$ is permitted to vary with $N$, even though this relationship is suppressed in the notation; see Remark \ref{rmk:changingweights} for further comments.
\end{remark}

\begin{remark} \label{rmk:clt}
Because $\Psi_s(\theta_0)$ is an average, $Z_s$ will be a zero-mean normally-distributed random variable provided that the weights $w_i$ and observations $Y_i$ are not too dependent on too many of their peers; see Chapters 20, 21 and 25 of \citepref{Davidson2021}. This is the scenario usually encountered in practice.
\end{remark}

\begin{remark}
The assumption given in \eqref{eqn:donsker} says that if $\hat{\theta}_s$ is close to $\theta_0$ (as it will be for large $N$), then $\sqrt{N} ( \Psi_s(\hat{\theta}_s) - \Psi(\hat{\theta}_s) )$ will be close to $\sqrt{N} (\Psi_s(\theta_0) - \Psi(\theta_0))$. Said another way, we require that the random function $\theta \mapsto \sqrt{N} ( \Psi_s(\theta) - \Psi(\theta) )$ possess a kind of continuity at $\theta = \theta_0$ that does not break down as $N \to \infty$, and \textit{stochastic equicontinuity} suffices; see Chapter 22.3 of \citepref{Davidson2021}. The set of random variables obtainable by evaluating $w_i \psi(Y_i ; \theta)$ at a given $\theta$ is said to be a \textit{Donsker class} if their variances have an upper bound and $\theta \mapsto \sqrt{N} ( \Psi_s(\theta) - \Psi(\theta) )$ is stochastically equicontinuous; see Chapter 2.1 of \citepref{vanderVaart1996}. By Lemma 3.3.5 of the same, if $(w_i, Y_i)$ is i.i.d.\ and $w_i \psi(Y_i ; \theta)$ is Donsker, then \eqref{eqn:donsker} is satisfied. To obtain \eqref{eqn:donsker} while accommodating dependence, combine a functional central limit theorem from Chapter 31 of \citepref{Davidson2021} with Addendum 1.5.8 of \citepref{vanderVaart1996}.
\end{remark}

This next corollary is a central limit theorem for $\sqrt{N}(\hat{\theta}_s - \hat{\theta}_N)$ to converge in distribution as $N \to \infty$ under the joint distribution spanning both the design and superpopulation. The asymptotic distribution is usually normal with zero mean. In the theorem after that, we will show that for design-unbiased weights, the joint variance of $\hat{\theta}_s - \hat{\theta}_N$ is close to the expected design variance of $\hat{\theta}_s$ if $N$ is large. As a result, this central limit theorem provides expressions for the asymptotic design variance that can be used to develop design-based variance estimators for $\hat{\theta}_s$.

\begin{corollary} \label{crl:normpop}
Suppose that the assumptions of Corollary \ref{crl:conspop} and Theorem \ref{thm:normsup} are satisfied for weights $w_1, w_2, \ldots, w_N$ and $1, 1, \ldots, 1$. Then we have
\begin{equation}
\sqrt{N}\dot{\Psi}_{\theta_0} (\hat{\theta}_s - \hat{\theta}_N) = - \sqrt{N}(\Psi_s(\theta_0) - \Psi_N(\theta_0)) + o_P(1). \label{eqn:linpop}
\end{equation}
If we also have that $\sqrt{N}(\Psi_s(\theta_0) - \Psi_N(\theta_0)) \Rightarrow Z^{\prime}$ for some distribution $Z^{\prime}$, then
\begin{equation}
\sqrt{N} (\hat{\theta}_s - \hat{\theta}_N) \Rightarrow \dot{\Psi}_{\theta_0}^{-1} Z^{\prime}. \label{eqn:convdistprime}
\end{equation}
\end{corollary}

\begin{remark}
Because $(\Psi_s(\theta_0), \Psi_N(\theta_0))$ is an average, it is typically asymptotically normal; see Remark \ref{rmk:clt}. Now apply the delta method (e.g.\ Theorem 3.1 of \citealppref{vanderVaart1998}) to obtain zero-mean asymptotic normality of $\sqrt{N}(\Psi_s(\theta_0) - \Psi_N(\theta_0))$.
\end{remark}

\subsection{Variance Estimation} \label{sec:varest}

The following theorem establishes the relationship between the design and joint variances of $\hat{\theta}_s$ for design-unbiased weights. We show in \eqref{eqn:designvar} and \eqref{eqn:designvartheta} that the joint variances of $\sqrt{N}(\Psi_s(\theta_0) - \Psi_N(\theta_0))$ and $\sqrt{N}(\hat{\theta}_s - \hat{\theta}_N)$ are equal and close to the expectation of their design-based counterparts, respectively. Combined with the central limit theorem in Corollary \ref{crl:normpop} above, this provides expressions for the asymptotic design variance that can be used to develop design-based variance estimators for $\hat{\theta}_s$, which is most useful for official statistics applications like survey design. More helpful for survey users conducting scientific inquiry, we also decompose in \eqref{eqn:uncondvar} and \eqref{eqn:uncondvartheta} the joint variance of $\sqrt{N}(\Psi_s(\theta_0) - \Psi(\theta_0))$ and $\sqrt{N}(\hat{\theta}_s - \theta_0)$, respectively, into a sum of: 1) the superpopulation expectation of their respective design variances, 2) the superpopulation variance of their respective population counterparts, and 3) a remainder term that is zero for $\Psi_s(\theta_0)$ and small for $\hat{\theta}_s$. This is helpful for scientific inquiry because it allows us to combine estimates of the design-based variance of $\hat{\theta}_s$ with estimates of the superpopulation variance of $\hat{\theta}_N$ to conduct statistical inference about $\theta_0$ as estimated by $\hat{\theta}_s$, taking into account uncertainty and variability from \textit{both} the survey design \textit{and} the study variables.

\begin{theorem} \label{thm:desvar}
Suppose that $\mathbb{E}[\Psi_s(\theta_0) \mid Y_{1:N}] = \Psi_N(\theta_0)$ for all $N$. Then
\begin{small}
\begin{align}
\mathrm{Var} \left( \sqrt{N}(\Psi_s(\theta_0) - \Psi_N(\theta_0)) \right) &= N \mathbb{E} \left[ \mathrm{Var} \left( \Psi_s(\theta_0) \mid Y_{1:N} \right) \right], \label{eqn:designvar} \\
\mathrm{Var} \left( \sqrt{N}(\Psi_s(\theta_0) - \Psi(\theta_0)) \right) &= N \mathbb{E} \left[ \mathrm{Var} \left( \Psi_s(\theta_0) \mid Y_{1:N} \right) \right] + N \mathrm{Var} \left( \Psi_N(\theta_0) \right). \label{eqn:uncondvar}
\end{align}
\end{small}
If we also satisfy the assumptions of Corollary \ref{crl:normpop} such that the second moment of every element of the $o_P(1)$ remainder in \eqref{eqn:linsup} converges to zero for weights $w_1, w_2, \ldots, w_N$ and $1, 1, \ldots, 1$, then
\begin{equation}
\mathrm{Var} \left( \sqrt{N}(\hat{\theta}_s - \hat{\theta}_N) \right) = N \mathbb{E} \left[ \mathrm{Var}(\hat{\theta}_s \mid Y_{1:N}) \right] + o(1). \label{eqn:designvartheta}
\end{equation}
Also assuming that the square of every element of $\sqrt{N}(\Psi_N(\theta_0) - \Psi(\theta_0))$ is asymptotically uniformly integrable (in the sense of Chapter 2.5 of \citealppref{vanderVaart1998}) gives
\begin{equation}
\mathrm{Var} \left( \sqrt{N}(\hat{\theta}_s - \theta_0) \right) = N \mathbb{E} \left[ \mathrm{Var}(\hat{\theta}_s \mid Y_{1:N}) \right] + \mathrm{Var} \left( \sqrt{N}(\hat{\theta}_N - \theta_0) \right) + o(1). \label{eqn:uncondvartheta}
\end{equation}
\end{theorem}

In view of \eqref{eqn:convdistprime}, \eqref{eqn:designvar} and \eqref{eqn:designvartheta}, the asymptotic design variance is given by
\begin{equation}
N \mathbb{E} \left[ \mathrm{Var}(\hat{\theta}_s \mid Y_{1:N}) \right] \to \dot{\Psi}_{\theta_0}^{-1} V^{\prime} (\dot{\Psi}_{\theta_0}^{-1})^{T}, \label{eqn:asydesignvar}
\end{equation}
where $V^{\prime}$ is the asymptotic variance of $\sqrt{N}(\Psi_s(\theta_0) - \Psi_N(\theta_0))$. Further, in view of \eqref{eqn:zestlin}, \eqref{eqn:uncondvar} and \eqref{eqn:uncondvartheta}, the asymptotic joint variance is given by
\begin{equation}
N \mathrm{Var}(\hat{\theta}_s) \to \dot{\Psi}_{\theta_0}^{-1} V_s (\dot{\Psi}_{\theta_0}^{-1})^T = \dot{\Psi}_{\theta_0}^{-1} (V^{\prime} + V) (\dot{\Psi}_{\theta_0}^{-1})^T, \label{eqn:asyuncondvar}
\end{equation}
where $V_s$ is the asymptotic variance of $\sqrt{N}(\Psi_s(\theta_0) - \Psi(\theta_0))$ and $V$ is the asymptotic variance of $\sqrt{N}(\Psi_N(\theta_0) - \Psi(\theta_0))$.

Given consistent estimators $\hat{\dot{\Psi}}_{\theta_0} \overset{P}{\to} \dot{\Psi}_{\theta_0}$ and $\hat{V}^{\prime} \equiv \widehat{\mathrm{Var}}(\sqrt{N} \Psi_s(\theta_0) \mid Y_{1:N}) \overset{P}{\to} V^{\prime}$, the continuous mapping theorem alongside \eqref{eqn:asydesignvar} suggests the following design-based variance estimator for $\hat{\theta}_s$:
\begin{equation}
\widehat{\mathrm{Var}}(\hat{\theta}_s \mid Y_{1:N}) = \frac{1}{N} \hat{\dot{\Psi}}_{\theta_0}^{-1} \hat{V}^{\prime} (\hat{\dot{\Psi}}_{\theta_0}^{-1})^{T}. \label{eqn:desvarest}
\end{equation}
In most (but not all) cases, $\dot{\Psi}_{\theta_0}$ can be consistently estimated by
\begin{equation}
\hat{\dot{\Psi}}_{\theta_0} = \frac{1}{N} \sum_{i = 1}^N w_i \dot{\psi}(Y_i ; \hat{\theta}_s), \label{eqn:dotpsiest}
\end{equation}
where the $\dot{\psi}(y ; \theta)$ is the Jacobian matrix of $\dot{\psi}(y ; \theta)$ with respect to $\theta$. This usually works if the Jacobian matrix of $\mathbb{E}[\psi(Y_i ; \theta)]$ equals $\mathbb{E}[\dot{\psi}(Y_i; \theta)]$ (i.e.\ if we can `move the derivative inside the expectation'). Quantiles are one notable case where this condition fails; see Section \ref{eg:quant}. In contrast to $\hat{\dot{\Psi}}_{\theta_0}$, expressions for $\hat{V}^{\prime}$ usually vary considerably depending on the sample design used. Fortunately, these expressions are often easily obtained using standard methods for producing design-based variance estimates of the population mean, but applied to $\psi(Y_i ; \hat{\theta}_s)$ in lieu of $Y_i$; see Section \ref{eg:sampling}.

If we also have a consistent estimator $\hat{V} \equiv \widehat{\mathrm{Var}}(\sqrt{N} \Psi_N(\theta_0)) \overset{P}{\to} V$, then the continuous mapping theorem alongside \eqref{eqn:asyuncondvar} suggests the following estimator for the joint variance of $\hat{\theta}_s$:
\begin{equation}
\widehat{\mathrm{Var}}(\hat{\theta}_s) = \frac{1}{N} \hat{\dot{\Psi}}_{\theta_0}^{-1} (\hat{V}^{\prime} + \hat{V}) (\hat{\dot{\Psi}}_{\theta_0}^{-1})^{T}. \label{eqn:uncondvaresttheta}
\end{equation}
If the $Y_i$ are i.i.d., then we can use the following unconditional variance estimator for $\sqrt{N} \Psi_N(\theta_0)$:
\begin{equation}
\hat{V} = \frac{1}{N} \sum_{i=1}^N w_i \psi(Y_i ; \hat{\theta}_s) \psi(Y_i ; \hat{\theta}_s)^T. \label{eqn:uncondvarest}
\end{equation}

\subsection{Normalised Weights} \label{sec:normweights}

We demonstrate here that for some estimators, normalising the weights before estimation has no asymptotic effect.

\begin{theorem} \label{thm:normweights}
For given weights $w_1, w_2, \ldots, w_N$, define the normalised weights $w^{\prime}_1, w^{\prime}_2, \ldots, w^{\prime}_N$ by
\begin{equation*}
w^{\prime}_i = \frac{w_i}{\frac{1}{N} \sum_{j=1}^N w_j},
\end{equation*}
and suppose that $\frac{1}{N} \sum_{j=1}^N w_j \overset{P}{\to} 1$. Then for an arbitrary estimator $\hat{\vartheta}_N$ and a fixed sequence $r_N$ the following are equivalent:
\begin{align}
\frac{1}{N} \sum_{i=1}^N w_i \psi(Y_i; \hat{\vartheta}_N) &= o_P(r_N), \label{eqn:normweightspsi}\\
\frac{1}{N} \sum_{i=1}^N w^{\prime}_i \psi(Y_i; \hat{\vartheta}_N) &= o_P(r_N). \label{eqn:unnormweightspsi}
\end{align}
\end{theorem}

\begin{remark}
Consider two estimators $\hat{\theta}_s$ and $\hat{\theta}^{\prime}_s$ that are constructed with weights $w_i$ and $w^{\prime}_i$ and satisfy \eqref{eqn:normweightspsi} and \eqref{eqn:unnormweightspsi}, respectively. This applies, for example, when producing linear regression coefficients (see Section \ref{eg:lr}) and maximum likelihood estimators (see Section \ref{eg:mle}). Theorem \ref{thm:normweights} implies that for both estimators, we can choose either $w_i$ or $w^{\prime}_i$ to establish consistency using Theorem \ref{thm:cons} (if $r_N \leq 1$), or to derive the asymptotic distribution using Theorem \ref{thm:normsup} and conduct variance estimation based on Theorem \ref{thm:desvar} (if $r_N \leq N^{-1/2}$). If the relevant assumptions are met, both estimators are consistent (if $r_N \leq 1$) or have the same asymptotic distributions and variances (if $r_N \leq N^{-1/2}$).
\end{remark}

\begin{remark}
Theorem \ref{thm:normweights} does not preclude the existence of estimators that satisfy \eqref{eqn:normweightspsi} and \eqref{eqn:unnormweightspsi} when constructed with normalised weights, but satisfy neither when constructed with unnormalised weights (or vice versa). The leading example is the quantile, which should be constructed with normalised weights and is covered in Section \ref{eg:quant}.
\end{remark}

\if0\wholepaper {
	\bibliographypref{../library}
} \fi

%% file: examples/examples.tex
\section{Examples} \label{sec:eg}

In the following examples we will illustrate how to produce an estimator $\hat{V}^{\prime}$ for the asymptotic design variance of $\sqrt{N} \Psi_s(\theta_0)$. Substituting $\hat{V}^{\prime}$ and \eqref{eqn:dotpsiest} into \eqref{eqn:desvarest} often gives a suitable estimator for the design variance of $\hat{\theta}_s$ that can be used for survey design. For quantiles (see Section \ref{eg:quant}), \eqref{eqn:dotpsiest} provides an inconsistent estimator $\hat{\dot{\Psi}}_{\theta_0}$ of $\dot{\Psi}_{\theta_0}$, and in this case we illustrate how a consistent $\hat{\dot{\Psi}}_{\theta_0}$ can be obtained. If the $Y_i$ are i.i.d., substitute \eqref{eqn:uncondvarest}, $\hat{\dot{\Psi}}_{\theta_0}$ and $\hat{V}^{\prime}$ into \eqref{eqn:uncondvaresttheta} to give a joint variance estimator for $\hat{\theta}_s$. When $Y_i$ is not independent, consistent estimators $\hat{V}$ for the asymptotic joint variance of $\sqrt{N} \Psi_N(\theta_0)$ can be used in lieu of \eqref{eqn:uncondvarest}; see Chapters 20, 21 and 25 of \citepref{Davidson2021}. Sections \ref{eg:weights} on weights and \ref{eg:sampling} on sampling leave the parameter $\theta_0$, its estimator $\hat{\theta}_s$ and the estimating equation $\Psi_s$ unspecified to highlight the generality of our approach, which is illustrated on specific statistics in Section \ref{eg:stats}.

\subsection{Weights} \label{eg:weights}

\subsubsection{Horvitz-Thompson} \label{eg:htweights}

Suppose that observations are obtained via a probability sample. In addition to the notation defined in Section \ref{sec:intest}, let the survey design's second-order selection probabilities be denoted by $\pi_{ij} = \mathrm{Pr}(\alpha_i = 1 \cap \alpha_j = 1)$, and let $A = \{i : \alpha_i = 1\}$ be the set of indices that identify the units in the probability sample. Since the Horvitz-Thompson weights $w_i = \alpha_i \pi_i^{-1}$ are design-unbiased (see \eqref{eqn:htunbiased} and surrounding discussion), Corollary \ref{crl:normpop} and Theorem \ref{thm:desvar} can be applied. The standard Horvitz-Thompson variance estimator of $\sqrt{N} \Psi_s(\theta_0)$ is given by (e.g.\ Result 2.8.1 of \citealppref{Sarndal1992})
\begin{equation}
\hat{V}^{\prime} = \frac{1}{N} \sum_{i \in A} \sum_{j \in A} (\pi_i^{-1} \pi_j^{-1} - \pi_{ij}^{-1}) \psi(Y_i ; \hat{\theta}_s) \psi(Y_i ; \hat{\theta}_s)^{T}. \label{eqn:htv}
\end{equation}
This variance estimator is approximately design unbiased if $\pi_{ij}$ is bounded away from zero for all $i,j$. If the sample size $n = \sum_{i=1}^N \alpha_i$ is not random, then the Yates-Grundy-Sen variance estimator for $\sqrt{N} \Psi_s(\theta)$ can be used instead; see Result 2.8.2 of \citepref{Sarndal1992}. Another alternative is the Hartley-Rao variance estimator \citeppref{Hartley1962}, which applies if without-replacement sampling is used.

\subsubsection{Data Integration} \label{eg:diweights}

Suppose our population is observed via a probability sample and a big-data set. If observations in the survey are accompanied by weights $w_i$ satisfying $\mathbb{E}[w_i \mid Y_{1:N}] = 1$, the integrated weights $w^{DI}_i = \delta_i + (1 - \delta_i) w_i$ are design-unbiased after extending $Y_{i}$ if needed so that its last element is equal to $\delta_i$ (see \eqref{eqn:diunbiased} and surrounding discussion). As a result, Corollary \ref{crl:normpop} and Theorem \ref{thm:desvar} can be applied.

Given that
\begin{align*}
\mathrm{Var} & (\sqrt{N} \Psi^{DI}_s(\theta_0) \mid Y_{1:N}) \\
&= \mathrm{Var} \left( \frac{1}{\sqrt{N}} \sum_{i=1}^N \delta_i \psi(Y_i ; \theta_0) + \frac{1}{\sqrt{N}} \sum_{i=1}^N (1 - \delta_i) w_i \psi(Y_i ; \theta_0)\ \Bigg\vert\ Y_{1:N} \right) \\
&= \mathrm{Var} \left( \frac{1}{\sqrt{N}} \sum_{i=1}^N w_i (1 - \delta_i) \psi(Y_i ; \theta_0)\ \Bigg\vert\ Y_{1:N} \right),
\end{align*}
we see that the design variance $\mathrm{Var}(\sqrt{N} \Psi^{DI}_s(\theta_0) \mid Y_{1:N})$ is equal to the variance of the survey-only estimate of $\frac{1}{\sqrt{N}} \sum_{i=1}^N (1 - \delta_i) \psi(Y_i ; \theta_0)$. Since this sum extends over only those observations not in the big data sample (or else the summand is zero), an integrated variance estimator is unavailable and we resort to the survey-only variance estimator. For example, if the survey weights $w_i$ are of the Horvitz-Thompson sort described in the previous subsection, then we obtain
\begin{equation}
\hat{V}^{\prime} = \frac{1}{N} \sum_{i \in A \setminus B} \sum_{j \in A \setminus B} (\pi_i^{-1} \pi_j^{-1} - \pi_{ij}^{-1}) \psi(Y_i ; \hat{\theta}^{DI}_s) \psi(Y_i ; \hat{\theta}^{DI}_s)^{T}, \label{eqn:ktv}
\end{equation}
where $B = \{i : \delta_i = 1 \}$ is the set of indices that identify the big-data observations.

In view of Remark \ref{rmk:clt}, both integrated and survey-only estimators are asymptotically unbiased by \eqref{eqn:zestlin} and \eqref{eqn:convdistprime}. One would therefore only choose the integrated estimator if it had a variance less than or equal to its survey-only counterpart. By Corollary \ref{crl:conspop} $\dot{\Psi}_{\theta_0}$ is identical for both integrated and Horvitz-Thompson estimation, so for scalar $\theta$ and beginning with the asymptotic design variance in \eqref{eqn:asydesignvar}, we would choose the integrated estimator if $\mathrm{Var}(\sqrt{N} \Psi^{DI}_s(\theta_0) \mid Y_{1:N}) \leq \mathrm{Var}(\sqrt{N} \Psi_s(\theta_0) \mid Y_{1:N})$. While this is difficult to show in general, if the survey weights are i.i.d.\ with nonzero variance outside a nonempty big-data set, then
\begin{align*}
\mathrm{Var}(\sqrt{N} \Psi^{DI}_s(\theta_0) \mid Y_{1:N}) &= \frac{1}{N} \sum_{i=1}^N (1 - \delta_i) \mathrm{Var}(w_i \mid Y_{1:N}) Y_i \\
&< \frac{1}{N} \sum_{i=1}^N \mathrm{Var}(w_i \mid Y_{1:N}) Y_i \\
&= \mathrm{Var}(\sqrt{N} \Psi_s(\theta_0) \mid Y_{1:N}).
\end{align*}
Underperformance of the integrated estimator could therefore only materialise if there were too much dependence between survey weights; if the covariance between different survey weights is small then the integrated estimator $\hat{\theta}^{DI}_s$ is more accurate than its survey-only counterpart $\hat{\theta}_s$ provided that the sample sizes (and thus the population size) are large enough. By Theorem \ref{thm:desvar}, an integrated estimator with a lower asymptotic design variance than its survey-only counterpart is also in possession of a lower asymptotic joint variance.

\subsection{Sampling} \label{eg:sampling}

\subsubsection{Simple Random Sampling Without Replacement} \label{eg:srswor}

Suppose that the population is sampled by taking a random draw of size $n$, without replacement. Assume that the sample size $n$ is obtained by rounding $fN$ to the nearest integer, where $f$ is a fixed sampling fraction, so that $\lvert n/N - f \rvert \leq 1/N$ and $n/N = f + O_P(N^{-1})$. Thus the asymptotic results above occur as both the sample size $n$ and the population size $N$ approach infinity in the rough proportion $n/N \approx f$ for a constant $f$. For a given population and sample we will use the asymptotic approximations that correspond to the sampling fraction $f = n/N$.

When constructing the integrated weights of \citepref{Kim2021} using Horvitz-Thompson survey weights, \eqref{eqn:ktv} becomes
\begin{gather}
\resizebox{0.9 \textwidth}{!}{$S^2 = \frac{1}{n-1} \left( \sum_{i \in A \setminus B} \psi(Y_i ; \hat{\theta}^{DI}_s) \psi(Y_i ; \hat{\theta}^{DI}_s )^{T} - \frac{1}{n} \left( \sum_{i \in A \setminus B} \psi(Y_i ; \hat{\theta}^{DI}_s) \right) \left( \sum_{i \in A \setminus B} \psi(Y_i ; \hat{\theta}^{DI}_s) \right)^{T} \right),$} \label{eqn:srswors2} \\
\hat{V}^{\prime} = \frac{1-f}{f} S^2, \label{eqn:srsworv}
\end{gather}
where we recall that $n$ is the sample size of $A$, the probability sample. Regarding the survey-only Horvitz-Thompson case of Example \ref{eg:htweights}, \eqref{eqn:htv} also becomes the above after setting the big-data sample $B$ to the empty set.

\subsubsection{Stratified Sampling} \label{eg:strat}

Consider a population of size $N$ comprised of $H$ strata indexed by $h = 1, 2, \ldots, H$, with each stratum $h$ containing a subpopulation $Y_{h,1}, Y_{h,2}, \ldots, Y_{h,N_h}$, such that $N = \sum_{h=1}^H N_h$. For each stratum $h$, the population size $N_h$ is obtained by rounding $F_h N$ to the nearest integer, where $F_h$ is a fixed `subpopulation fraction'. As a result, $\lvert N_h / N - F_h \rvert \leq 1/N$ and $N_h / N = F_h + O_P(N^{-1})$. Thus the asymptotic results above occur as both the subpopulation sizes $N_h$ and the population size $N$ approach infinity according to the rough proportions $N_h/N \approx F_h$ for constants $F_h$. For a given stratification we will use the asymptotic approximations that correspond to the subpopulation fractions $F_h = N_h / N$. Supposing that each variable $Y_{h,i}$ is assigned the weight $w_{h,i}$, we can apply the results of Section \ref{sec:theory} via reindexation, for example by imposing $Y_{\sum_{k=1}^{h-1} N_k + i} = Y_{h,i}$ and $w_{\sum_{k=1}^{h-1} N_k + i} = w_{h,i}$. This has the effect of making averages across $(h,i)$ equivalent to averages across $i$ when applying the results of Section \ref{sec:theory}.

Assume that the strata are sampled independently given $Y_{1:N}$, and let $\Psi_h(\theta) = \frac{1}{N_h} \sum_{i=1}^{N_h} w_{h,i} \psi(Y_{h,i}; \theta)$. Then
\begin{align*}
\Psi_s(\theta) &= \sum_{h=1}^H F_h \Psi_h(\theta), \\
\mathrm{Var}(\sqrt{N} \Psi_s(\theta) \mid Y_{1:N}) &= \sum_{h=1}^H F_h \mathrm{Var}(\sqrt{N_h} \Psi_h(\theta) \mid Y_{1:N}).
\end{align*}
Given estimators $\hat{V}^{\prime}_h$ for $\mathrm{Var}(\sqrt{N_h} \Psi_h(\theta_0) \mid Y_{1:N})$, an estimator $\hat{V}^{\prime}$ for $\mathrm{Var}(\sqrt{N} \Psi_s(\theta_0) \mid Y_{1:N})$ is therefore given by
\begin{equation}
\hat{V}^{\prime} = \sum_{h=1}^H F_h \hat{V}^{\prime}_h. \label{eqn:vhatprimestrat}
\end{equation}

Suppose that we produce a sample of size $n_h$ from each stratum $h$ independently, using simple random sampling without replacement to observe each $Y_{h,i}$ whose index $i$ lies in a set $A_h$. Also suppose that for each stratum $h$ there is a set of indices $B_h$ identifying those observations $Y_{h,i}, i \in B_h$ that are obtained via the big data sample. If we use \citepref{Kim2021} integrated weights constructed with Horvitz-Thompson survey weights, then \eqref{eqn:srswors2} and \eqref{eqn:srsworv} can be applied within each stratum to give
\begin{gather}
f_h = \frac{n_h}{N_h}, \nonumber \\
\resizebox{0.9 \textwidth}{!}{$S^2_h = \frac{1}{n_h-1} \left( \sum_{i \in A_h \setminus B_h} \psi(Y_i ; \hat{\theta}^{DI}_s) \psi(Y_{h,i} ; \hat{\theta}^{DI}_s )^{T} - \frac{1}{n_h} \left( \sum_{i \in A_h \setminus B_h} \psi(Y_{h,i} ; \hat{\theta}^{DI}_s) \right) \left( \sum_{i \in A_h \setminus B_h} \psi(Y_{h,i} ; \hat{\theta}^{DI}_s) \right)^{T} \right),$} \nonumber \\
\hat{V}^{\prime}_h = \frac{1 - f_h}{f_h} S_h^2. \label{eqn:vhatprimestratintegratedsrswor}
\end{gather}

If there is no big-data set, the above applies after setting $B_h = \varnothing$. Following \citepref{Lohr2021}, another leading case is where the big data sample is treated as a completely enumerated stratum, whereby $A_1 = \varnothing$ and $B_h = \varnothing$ for all $h = 2, 3, \ldots, H$, say. In this case, we survey only those units not in the big data set. For a given survey sample of size $n$ this strategy is more efficient than surveying the entire population. For example, if $H = 2$, it is not too difficult to show using \eqref{eqn:asydesignvar} that the asymptotic design variance obtained by surveying only non-big-data units is $(F_2^2 - f)/(1 - f)$ times the variance obtained by sampling from the entire population, where $f = \sum_{h=1}^H f_h$ is the total sampling fraction. If $\hat{\theta}^{DI}_s$ is a scalar estimator, we can optimally allocate the total sampling fraction to strata by using standard constrained optimisation techniques to minimise \eqref{eqn:desvarest} across $f_1, f_2, \ldots, f_H$ subject to $\sum_{h=1}^H f_h = f$ and $0 \leq f_h \leq 1, h = 1, 2, \ldots, H$.

\subsection{Statistics} \label{eg:stats}

In defining the statistics below, we will sometimes make use of the following estimate for the cumulative distribution function (c.d.f.):
\begin{equation*}
\hat{F}_s(y) = \frac{1}{N} \sum_{i=1}^N w_i I(Y_i \leq y).
\end{equation*}
Note that $\hat{F}_s(y)$ is not necessarily a c.d.f.\ itself, unless $w_i \geq 0$ and $\frac{1}{N} \sum_{i=1}^N w_i = 1$. When assuming that the population $Y_1, Y_2, \ldots, Y_N$ is i.i.d., we will use $F$ and $f$ to denote the c.d.f.\ and density, respectively, of $Y_i$. 

\subsubsection{Quantiles} \label{eg:quant}

Let $Y_{(i)}$ be the $i$\textsuperscript{th} smallest value in the population, so that $Y_{(1)} \leq Y_{(2)} \leq \cdots \leq Y_{(N)}$, and let $w_{(i)}$ be the weight associated with $Y_{(i)}$. Consider the $p$-quantile given by
\begin{equation}
\hat{\theta}_s = \inf \{ y : \hat{F}_s(y) \geq p \}, \label{eqn:pquantile}
\end{equation}
and the function
\begin{equation*}
\psi(y ; \theta) = (1 - p) I(y < \theta) - p I(y > \theta).
\end{equation*}
By definition, there exists a $j$ such that $\hat{\theta}_s = Y_{(j)}$ and $p \leq \hat{F}_s(\hat{\theta}_s) \leq p + w_{(j)}/N$. As a result, if $\frac{1}{N} \sum_{i=1}^N w_{i} = 1$ and $Y_i$ is continuously distributed (so that $\frac{1}{N} \sum_{i=1}^N w_{i} I(Y_{i} = \hat{\theta}_s) = w_{(j)}/N$), then we have
\begin{equation*}
-(1-p) w_{(j)} \leq N \Psi_s(\hat{\theta}_s) \leq p w_{(j)}.
\end{equation*}
Integrability of $w_{(j)}$ (implied by boundedness for example) then provides $\Psi_s(\hat{\theta}_s) = o_P(N^{-1/2})$, as required by Theorem \ref{thm:normsup}. If the population is i.i.d., then $\Psi(\theta) = \mathbb{E}[\psi(Y_i ; \theta)] = F(\theta) - p$ and $\dot{\Psi}_{\theta_0} = f(\theta_0)$. Note that in this case, \eqref{eqn:dotpsiest} cannot be used to estimate $\dot{\Psi}_{\theta_0}$, since $\dot{\psi}(Y_i ; \hat{\theta}_s)$ is undefined if $Y_i = \hat{\theta}_s$. Given an estimator $\hat{f}$ for the density $f$ (e.g.\ \citealppref{Buskirk2005}), we can use $\hat{\dot{\Psi}}_{\theta_0} = \hat{f}(\theta_0)$ instead.

\subsubsection{Gini Index} \label{eg:gini}

Consider the the weighted Gini index given by
\begin{equation}
\hat{\theta}_s = \frac{\int \int \lvert y - x \rvert\ \mathrm{d} \hat{F}_s(x)\ \mathrm{d} \hat{F}_s(y)}{2 \int y\ \mathrm{d} \hat{F}_s(y)} = \frac{\sum_{i=1}^N \sum_{j=1}^N w_i w_j \lvert Y_i - Y_j \rvert}{2 N \sum_{i=1}^N w_i Y_i}. \label{eqn:gini}
\end{equation}
If the population is generated i.i.d., then we will show that the estimating equation following from
\begin{equation*}
\psi(y ; \theta) = 2 \int (I(y \leq x) - F(x)) x\ \mathrm{d} F(x) + (2 F(y) - 1)y - \theta y
\end{equation*}
satisfies $\Psi_s(\theta) = o_P(N^{-1/2})$, as required by Theorem \ref{thm:normsup}. Rearrangement of \eqref{eqn:gini} gives
\begin{equation*}
\hat{\theta}_s = \frac{\int (2 \hat{F}_s(y) - 1) y\ \mathrm{d} \hat{F}_s(y) - \frac{1}{N^2} \sum_{i=1}^N w_i^2 Y_i}{\int y\ \mathrm{d} \hat{F}_s(y)},
\end{equation*}
so that
\begin{equation*}
\sqrt{N} \Psi_s(\hat{\theta}_N) = - 2 \int (\hat{F}_s(y) - F(y)) y\ \mathrm{d} \left[ \sqrt{N} (\hat{F}_s(y) - F(y)) \right] + \frac{1}{\sqrt{N}} \frac{1}{N} \sum_{i=1}^N w_i^2 Y_i.
\end{equation*}
For i.i.d.\ $(Y_1, w_1), (Y_2, w_2), \ldots, (Y_N, w_N)$, the first term is $o_P(1)$ by Example 2.10.27,\footnote{Invoke with the upper bound $y \mapsto \max(y, 0)$. By symmetry obtain the same for nonpositive, nondecreasing functions with lower bound $y \mapsto \min(y, 0)$.} Example 2.10.7,\footnote{Invoke first with $\mathcal{F} = \{f : 0 \leq f(y) \leq \max(y, 0) \}$ and $\mathcal{G} = \{g : \min(y, 0) \leq g(y) \leq 0 \}$; second with $\{ y \mapsto f(y) + g(y) : f \in \mathcal{F}, g \in \mathcal{G} \}$ and $\{ y \mapsto -(f(y) + g(y)) : f \in \mathcal{F}, g \in \mathcal{G} \}$.} Example 2.10.10,\footnote{Let $\mathcal{F} = \{ (y,w) \mapsto f(y)y : -1 \leq f(y) \leq 1 \}$ and $g(y,w) = w$.} Equation 2.1.8 (with surrounding discussion) and  Problem 2.9.1 of \citepref{vanderVaart1996}, provided that the weights are almost surely bounded and $\mathbb{E}[Y_i^{2 + \epsilon}] < \infty$ for some $\epsilon > 0$. The second term converges to zero after applying a weak law of large numbers to $\frac{1}{N} \sum_{i=1}^N w_i^2 Y_i$.

A law of large numbers gives $\Psi_s(\theta) \overset{P}{\to} \Psi(\theta) = \mathbb{E}[\psi(Y_i; \theta)] = \mathbb{E}[(2 F(Y_i) - 1) Y_i] - \theta \mathbb{E}[Y_i]$ so that $\dot{\Psi}_{\theta_0} = - \mathbb{E}[Y_i] = \mathbb{E}[\dot{\psi}(Y_i ; \theta)]$, and \eqref{eqn:dotpsiest} suffices, giving $\hat{\dot{\Psi}}_{\theta_0} = - \frac{1}{N} \sum_{i=1}^N w_i Y_i$. When applying any of the above formulas for $\hat{V}^{\prime}$, use $\hat{\psi}(y ; \theta) = 2 \int (I(y \leq x) - \hat{F}_s(x))x\ \mathrm{d} \hat{F}_s(x) + (2 \hat{F}_s(y) - 1)y - \theta y$ in place of $\psi(y ; \theta)$, since $F$ is unknown.

\subsubsection{Linear Regression Coefficients} \label{eg:lr}

Suppose that we treat the first element $Y_{i,1}$ as a scalar regressand and $X_i = Y_{i,2:k}$ as a $(k-1)$-dimensional row vector of regressors. If $\hat{\theta}_s$ is the $(k-1)$-dimensional column vector of linear least-squares coefficients minimising
\begin{equation*}
\theta \mapsto \frac{1}{N} \sum_{i=1}^N w_i (Y_{i,1} - X_i \theta)^2,
\end{equation*}
then differentiating the above with respect to $\theta$ reveals that letting $\psi^{\prime}(y,x ; \theta) = x^T (y - x \theta)$ and $\psi(y ; \theta) = \psi^{\prime}(y_1, y_{2:k} ; \theta)$ gives $\Psi_s(\hat{\theta}_s) = 0 = o_P(N^{-1/2})$, as required by Theorem \ref{thm:normsup}. For an i.i.d.\ superpopulation with finite second moments, $\Psi(\theta) = \mathbb{E}[X_i^T Y_i] - \mathbb{E}[X_i^T X_i] \theta$ and $\dot{\Psi}_{\theta_0} = - \mathbb{E}[X_i^T X_i]$, which can be estimated by $\hat{\dot{\Psi}}_{\theta_0} = - \frac{1}{N} \sum_{i=1}^N w_i X_i^T X_i$. This is the same estimate given by \eqref{eqn:dotpsiest}.

When using Horvitz-Thompson weights, the above $\hat{\theta}_s$ contains the same estimated regression coefficients computed in Section 4.2 of \citepref{Binder1983} and Example 5.1 of \citepref{Pfeffermann1993}, and shown to be optimal in Theorem 1 of \citepref{Godambe1986}. Our asymptotic joint variance in \eqref{eqn:asyuncondvar} is new here, as is the joint variance estimator in \eqref{eqn:uncondvaresttheta}. \citepref{Pfeffermann1993} uses only the design variance on the basis that it is close to the joint variance when the population is much larger than the sample (see his Equation 2.4). In our view this further approximation is unnecessary because computing the joint variance estimator is almost as easy as computing the design-based variance estimator in \eqref{eqn:desvarest}, and it is not always true that the sample is small relative to the population.

\subsubsection{Maximum Likelihood Estimator} \label{eg:mle}

Suppose that we want to fit to the population a postulated (but not necessarily correctly specified) model parameterised by a vector $\theta$ with likelihood function $p(y ; \theta)$. Consider the estimator $\hat{\theta}_s$ that maximises the weighted likelihood $\frac{1}{N} \sum_{i=1}^N w_i \log p(Y_i ; \theta)$ across all $\theta$ in a parameter space. If $\hat{\theta}_s$ converges in probability to a fixed value $\theta_0$ in the interior of the parameter space and $\psi(y ; \theta)$ is the gradient of the weighted likelihood with respect to $\theta$, then $\Psi_s(\hat{\theta}_s)$ equals zero with probability approaching one and $\Psi_s(\hat{\theta}_s) = o_P(N^{-1/2})$ follows, as required by Theorem \ref{thm:normsup}.

\if0\wholepaper {
	\bibliographypref{../library}
} \fi

%% file: simulation/simulation.tex
\section{Simulation}\label{sec:sim}

\subsection{Overview}

In this simulation exercise, we study the performance of integrated estimation of the median and Gini index in the context of Australian personal income. We shall compare the bias and variance of big-data-only, survey-only and integrated estimators across varying population sizes, maintaining a fixed sampling fraction. Estimators are compared under the joint distribution spanning both the design and superpopulation. 

In this case, our population consists of $12$ strata stratified by age and sex (males and females aged 24 and under, 25 to 34, 35 to 44, 45 to 54, 55 to 64 and 65 and over), where each stratum is comprised of scalar observations representing personal income. Our big-data set will emulate an administrative tax-return data set as could be sourced from a taxation department such as the Australian Taxation Office. Because not all Australian residents earning under the tax-free threshold of \$18,200 are required to submit a tax-return, we expect that such a data set would underrepresent this demographic, introducing selection bias that will be emulated by the big-data sampling mechanism in the simulation. Our simulated survey data set will represent an official survey conducted by a national statistics office, such as the Australian Bureau of Statistics' Survey of Income and Housing (SIH), and use stratified simple random sampling without replacement.\footnote{Note that the SIH typically uses a survey design different from what we consider here; see \citepref{ABSMD}.} We consider two survey sampling mechanisms: one in which units are sampled from the entire population, and a second in which only non-big-data units are sampled. 

\subsection{Simulation Design}

Firstly, we construct a superpopulation distribution from which an `Australia-like' population may be generated. This superpopulation distribution is a finite mixture with c.d.f.
\begin{equation}
F(y) = \sum_{h = 1}^{12} p_h F_h(y),
\end{equation}
where $p_h$ represents the proportion of stratum $h$ in the Australian population, and $F_h(y)$ the c.d.f.\ of stratum $h$. These proportions are derived from Table 4.1 of \citepref{ABSPI} by calculating the ratio of the counts (Column I) of each stratum for a given age (Column C) and sex (Column D) to the total (Column I, Row 28). Each c.d.f.\ $F_h(y)$ is constructed by interpolating a monotonic cubic smoothing spline through income frequency data from Graph 1 of \citepref{ABSDB}, using the method described in the `Interpolated CDFs' section of \citepref{Hippel2017}. Specifically, for each stratum $h$, we fit a cubic smoothing spline with knots $(y, F_h(y))$ at the points 
\begin{equation}
( 0, 0), \bigg ( \frac{52\eta_h b_1}{\eta}, \frac{ r_1}{100} \bigg ),  \bigg ( \frac{52\eta_h b_2}{\eta}, \frac{ \sum_{i=1}^2 r_i}{100} \bigg ), \ldots,  \bigg ( \frac{52\eta_h b_{57}}{\eta}, 1 \bigg ),
\end{equation}
where $\eta_h$ is the reported median of stratum $h$ (Table 4.1 Col.\ N of \citealppref{ABSPI}), $\eta$ is the reported population median (Table 4.1 Col.\ N of \citealppref{ABSPI}), $b_i$ is the upper bound of the $i$\textsuperscript{th} income bracket ($x$-axis) of Graph 1 of \citepref{ABSDB}\footnote{Note that \citepref{ABSDB} uses equivalised household income instead of personal income. We assume that the distribution is comparable. } and $r_i$ is the corresponding frequency from the 2019-20 financial year ($y$-axis). Bracket $b_{57}$ is set such that $\mathbb{E}[Y_h]= \mu_h$, where $\mu_h$ is the reported mean of stratum $h$ (Table 4.1 Col.\ S of \citealppref{ABSPI}). This approach also ensures the median of each stratum $h$ falls within the income bracket of the reported median $\eta_h$. The coefficients of each spline are computed by the Hyman method using the splinebins function of the R package `Binsmooth' (\citealppref{binsmooth}).

From this c.d.f.\ $F$, we calculate the superpopulation median, given by $F^{-1}(0.5)$, and the superpopulation Gini index, given by $\mathbb{E}[Y]^{-1}\int_0^\infty F(y)(1 - F(y)) \mathop{dy}$. We also estimate the asymptotic joint variance (see Theorem \ref{thm:desvar}) of the integrated median and Gini index estimators by generating $10^8$ observations from the superpopulation distribution and using these observations to calculate $\hat V ^\prime$ based on (\ref{eqn:vhatprimestrat}) and (\ref{eqn:vhatprimestratintegratedsrswor}).

We then perform a simulation consisting of the following steps:
\begin{enumerate}
	\item Generate $2 \times 10^6$ observations from the superpopulation distribution, given above.
	\item Construct $10$ populations of increasing sizes, the sizes equally spaced along the interval $[ 5 \times 10^5,  \ldots, 2 \times 10^6]$, where the population $U_k = \{1, 2, \ldots, N_k\}$ consists of the first $N_k$ individuals whose incomes were generated in Step 1. Denote $Y_{i}$ the income of the $i$\textsuperscript{th} individual.
	\item For each population $U_k$, sample $N_k/2$ individuals to comprise big data set $B_k$, where the probability of individual $i \in U_k$ being selected into $B_k$ is given by
	\begin {equation}
	\pi_i^b =
    \begin{cases}
        (|H_k| + 0.05|L_k| )^{-1} &  i \in H_k\\
        0.05(|H_k| + 0.05|L_k|)^{-1} &  i \in L_k
    \end{cases},
    \end{equation}
    where $ L_k = \{i \in U_k  \mid Y_i < 18200 \} $ and $ H_k = U_k \setminus L_k$. Note that we  $\pi_i^b$ would be unknown in practice and is not used to compute any statistics in this simulation.
	\item For each population $U_k$ and stratum $h$, sample
	\begin{equation}
	n_{k, h} =  n_k \frac{N_{k,h} S_{Y_{k,h}}}{\sum_{h=1}^{12} N_{k, h} S_{Y_{k, h}}}
	\end{equation}
	units without replacement to produce the set of sampled stratum-$h$ units units $A_{k,h}$ and the set of all surveyed units $A_k = \cup_{h=1}^{12} A_{k,h}$,  where $N_{k, h}$ is the population size of stratum $h$ from population $U_k$, $n_k = 10^{-3}N_k$,  and $S_{Y_{k, h}}^2$ is the sample variance of $Y_i$ across $i$ in stratum $h$ and population $U_k$. Within each stratum, units are selected with equal probability, yielding a first-order inclusion probability of $ \pi_i = n_{k,h}/N_{k, h}$ for a unit $i$ in stratum $h$. Given a fixed total sample size $n_{k}$, this allocation minimises the variance of the survey-only mean (see Section 3.7.4.\ i.\ of \citealppref{Sarndal1992}).
	\item Define $ U_k^{\prime} = U_k \setminus B_k$. Repeat Step 4., sampling instead from $U_k^{\prime}$ to form $A_k^{\prime}$. This is equivalent to treating the big data sample as a completely enumerated stratum, as described in Section \ref{eg:strat}.
	\item For $k \in \{1,2,\ldots,10\}$, use \eqref{eqn:pquantile} with $p = 0.5$ to calculate the following estimates of the median:
	\begin{itemize}
		\item An unweighted estimate using only $B_k$. 
		\item A Horvitz-Thompson-weighted estimate using only $A_k$.
		\item An integrated estimate using both $B_k$ and $A_k$, using the integrated weights in \eqref{eqn:wdi} alongside Horvitz-Thompson survey weights.
		\item An integrated estimate using both $B_k$ and $A_k^{\prime}$, using the integrated weights in \eqref{eqn:wdi} alongside Horvitz-Thompson survey weights.
	\end{itemize}
	\item Repeat Step 6.\ for the Gini index given by $\eqref{eqn:gini}$.
	\item Produce $10^4$ Monte-Carlo draws for each population size $N_k$ by repeating Steps 1 to 7.
	\item Across incomes given each $N_k$:
	\begin{enumerate}
	\item Calculate the sample variance of the four estimates for both the median and Gini index using the Monte-Carlo draws.
	\item Calculate the sample bias of the four estimates using the Monte-Carlo draws and the superpopulation median and Gini index. 
	\item Calculate approximate $95 \%$ confidence intervals for the bias and variance estimates, using the standard asymptotics for i.i.d.\ draws.
	\end{enumerate}
\end{enumerate}

\subsection{Results}

\begin{figure*}[p]
\centering
\includegraphics[height=0.4\textheight]{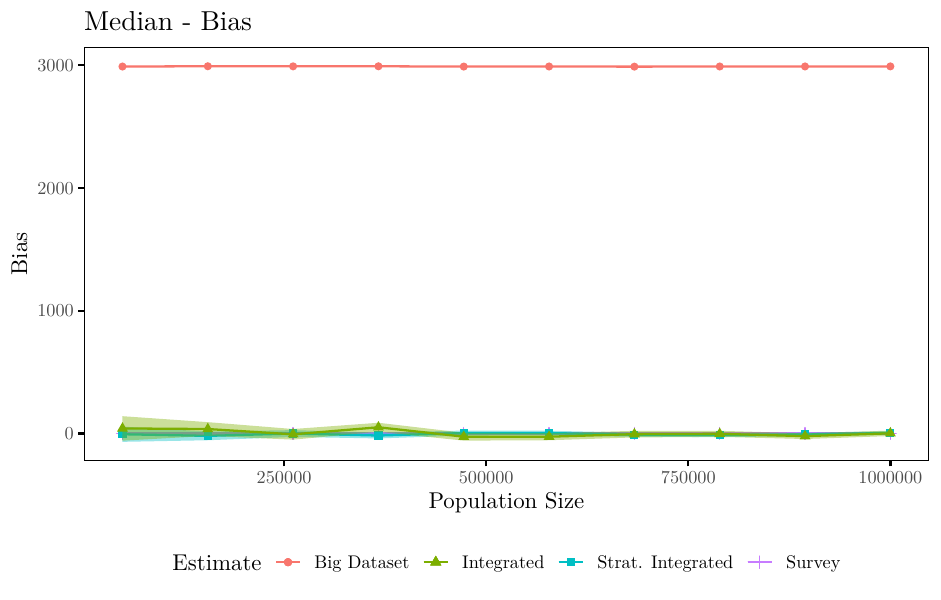}
\caption {\small Sample bias ($y$-axis) of estimates of the median (line type) for different population sizes ($x$-axis), with 95\% confidence intervals represented as shaded ribbons. Here, `Strat.\ Integrated' refers to the integrated estimate that uses $A^{\prime}$ (the survey sampled from non-big-data units only).}
\label{fig:median_bias}
\end{figure*}

\begin{figure*}[p]
\centering
\includegraphics[height=0.4\textheight]{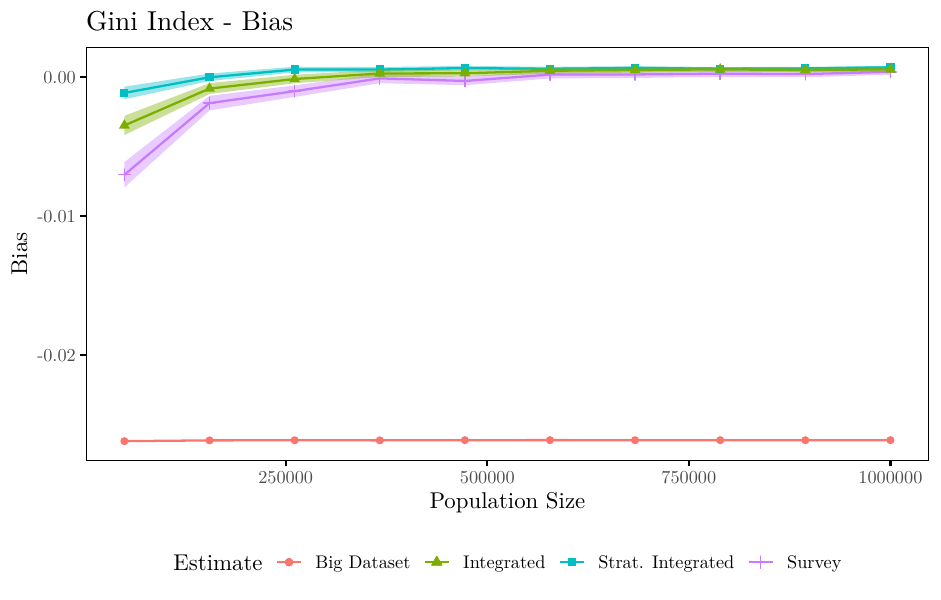}
\caption {\small Sample bias ($y$-axis) of estimates of the Gini index (line type) for different population sizes ($x$-axis), with 95\% confidence intervals represented as shaded ribbons. Here, `Strat.\ Integrated' refers to the integrated estimate that uses $A^{\prime}$ (the survey sampled from non-big-data units only).}
\label{fig:gini_bias}
\end{figure*}

\begin{figure*}[p]
\centering
\includegraphics[height=0.37\textheight]{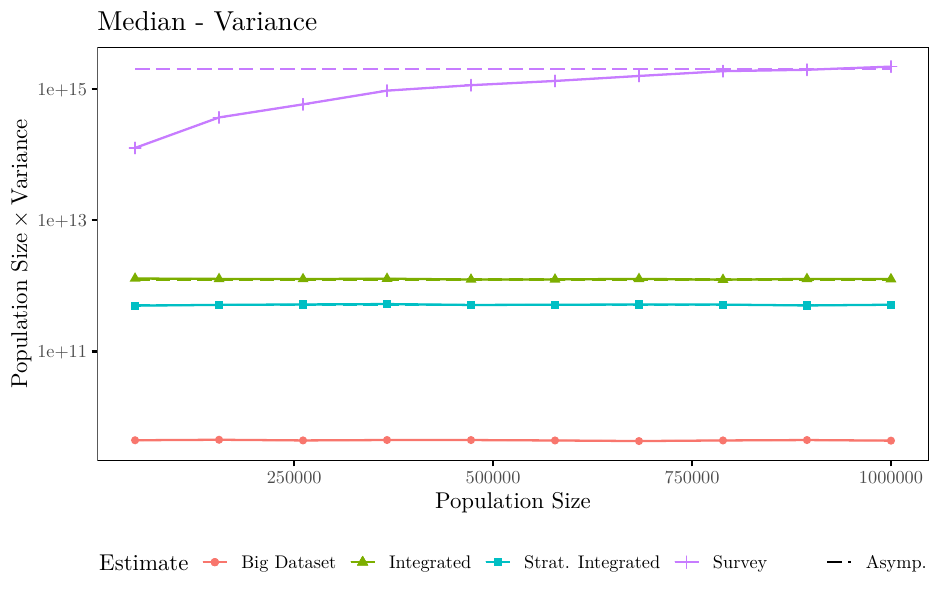}
\caption {\footnotesize Sample population-size-adjusted variance ($y$-axis) of estimates of the median (line type) on a $\log_{10}$ scale for different population sizes ($x$-axis). The corresponding 95\% confidence intervals are around the width of the lines, and have been omitted. Here, `Strat.\ Integrated' refers to the integrated estimate that uses $A^{\prime}$ (the survey sampled from non-big-data units only). The dashed lines depict the asymptotic variance of the survey-only and integrated estimators, given by Theorem \ref{thm:desvar}.}
\label{fig:median_var}
\end{figure*}

\begin{figure*}[p]
\centering
\includegraphics[height=0.37\textheight]{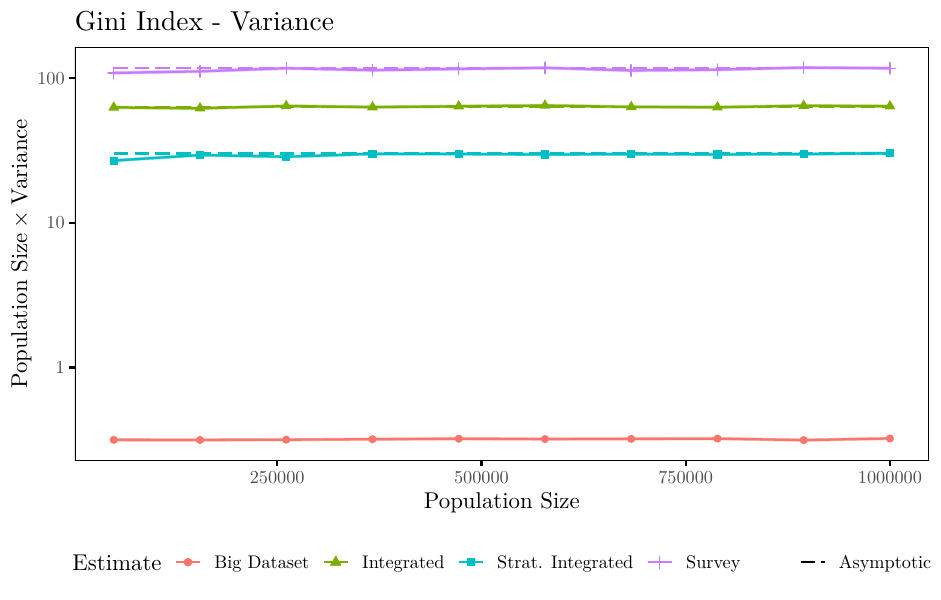}
\caption {\footnotesize Sample population-size-adjusted variance ($y$-axis) of estimates of the Gini index (line type) on a $\log_{10}$ scale for different population sizes ($x$-axis). The corresponding 95\% confidence intervals are around the width of the lines, and have been omitted. Here, `Strat.\ Integrated' refers to the integrated estimate that use $A^{\prime}$ (the survey sampled from non-big-data units only). The dashed lines depict the asymptotic variance of the survey-only and integrated estimators, given by Theorem \ref{thm:desvar}.}
\label{fig:gini_var}
\end{figure*}

The sample bias of the median and Gini index estimates as a function of the population size are depicted in Figures \ref{fig:median_bias} and \ref{fig:gini_bias} respectively. Observe that the big-data-only estimate exhibits bias due to the sampling mechanism of the big data set outlined in Step 3. This mechanism overrepresents high-income earners, resulting in an overestimate of the median and an underestimate of the Gini index. Importantly, both of the integrated estimators yield a statistically insignificant bias for large populations, which is consistent with Theorem \ref{thm:normsup} and Remark \ref{rmk:clt}.

Figures \ref{fig:median_var} and \ref{fig:gini_var} depict the size-adjusted variance estimates for the Gini index and median, respectively. We see that the Monte-Carlo variance of the integrated estimators calculated in Step 9.\ are consistent with the asymptotic variance estimates described in Section \ref{sec:varest}. This demonstrates the utility of the asymptotic variance estimates for practical applications, particularly given that the asymptotic variances only very occasionally fall outside the confidence intervals for the corresponding Monte-Carlo variances for population and sample sizes consistent with those of surveys such as the SIH.

Figures \ref{fig:median_var} and \ref{fig:gini_var} also highlight the efficiency of the integrated estimators over the survey-only estimators in that they exhibit significantly lower variance. Additionally, we can see from both Figure \ref{fig:median_var} and Figure \ref{fig:gini_var} that the integrated estimate using probability sample $A_k^{\prime}$ outperforms in terms of variance the estimate using $A_k$ (with a variance of less than half). This illustrates that, in the presence of a pre-existing big data set, significant efficiency gains in terms of variance can be achieved by tailoring the survey design to exclude those units present in the big data set.

\if0\wholepaper {
	\bibliographypref{../library}
} \fi

%% file: conclusion/conclusion.tex
\section{Conclusion} \label{sec:conclusion}

In this paper we have explored a general method for survey design and estimation in the presence of big data, in a way that accommodates many different weights, sampling approaches and statistics. Our method leads to an estimator that is asymptotically unbiased (both under the design and jointly with the superpopulation) and where both design-based and joint variance estimators are available. Using the method to integrate big and survey data reduces the design and joint variance of the estimator, provided that dependence between the variables of different units is small. All results are validated and illustrated in a simulation study that examines the performance of estimates of the median and Gini index in the case of Australian personal income.

Several directions of inquiry for future research are evident. First, note that our approach for combining big and survey data to produce nonlinear statistics has been explored theoretically and in simulation above, but the benefits of this approach in practice can only be precisely quantified by applying it to real-world data.

Second, recall that in Theorem \ref{thm:desvar} we assume that the weights are design unbiased in order to produce valid variance estimators, but many popular estimators that take advantage of auxiliary characteristics do not satisfy this property. The generalised regression estimator is a leading example, and is not design unbiased because the coefficients must be estimated. While such estimators may be asymptotically design unbiased, this is not likely to be sufficient because of the $N$ terms in \eqref{eqn:designvar} and \eqref{eqn:uncondvar}. One possible solution might be to incorporate into $\theta$ the coefficients relating the auxiliary characteristics to the target variable, and use \textit{two-step} estimation (e.g.\ Section 6 of \citealppref{Newey1994}) to recover the marginal asymptotic variance of the elements in $\hat{\theta}_s$ that are of interest. This might even extend to cases where models and auxiliary characteristics are used to correct for biases such as measurement error, or incorporate information from big data for which only uncertain linkage to the survey is available (e.g.\ \citealppref{Fellegi1969, Samuels2012}).

Third, we saw in \eqref{eqn:desvarest} and \eqref{eqn:dotpsiest} that variance estimation of nonlinear statistics often involves taking derivatives that can be tedious and error prone to do manually. As a result, existing software packages for survey design and variance estimation advance on their predecessors in part by making better use of automatic differentiation facilities to broaden the class of estimators supported by their variance estimation routines. For example, CLAN \citeppref{Andersson1994} advances on traditional computing environments (e.g.\ SAS) by supporting variance estimation for rational functions of linear estimators of totals. In turn, ReGenesees \citeppref{Zardetto2015} advances on CLAN by extending this support to functions whose derivatives may be computed via the `deriv' routine provided in base R \citeppref{R2020}. In this case, supported functions are formed by composing `$+$', `$-$', `$\times$', `$/$', `exp', `log', `sin', and several others. Such software packages (see references in \citealp{Zardetto2015}) predate recently developed automatic diffferentiation libraries (e.g.\ \citealppref{Carpenter2015, Paszke2017, Paszke2019}) that better support a greater range of functions. Further, in Section 9, \citepref{Zardetto2015} proposes exploring functional derivatives as a way of again broadening the class of estimators supported by variance estimation routines, and the estimating equations framework discussed in the current paper provides an alternative approach. We therefore contend that the time is ripe for the development of a new software library for survey design and variance estimation that takes advantage of these recent developments and supports a greater range of nonlinear statistics incorporating information from both surveys and big data.

\if0\wholepaper {
	\bibliographypref{../library}
} \fi

%% file: supplemental_data_title_page.tex
\singlespacing

\begin{titlingpage}

\if0\blind
{
	\renewcommand{\thefootnote}{\fnsymbol{footnote}}
	\title{\papertitle: Supplemental Data}
	\author{\paperauthors}
	\date{\mydate}
	\maketitle
	\footnotetext[1]{\ryanfootnote}
	\footnotetext[2]{\luccafootnote}
	\footnotetext[4]{\mddisclaimer}
	\setcounter{footnote}{0}
	\renewcommand{\thefootnote}{\arabic{footnote}}
} \fi

\end{titlingpage}

%% file: supplemental.tex
\renewcommand{\wholepaper}{1}

\doublespacing

\appendix

\setcounter{section}{0}

\subfile{proofs/proofs.tex}


%% file: proofs/proofs.tex
\section{Proofs}

\begin{proof}[Proof of Theorem \ref{thm:cons}]
Apply Theorem 5.9 of \citepref{vanderVaart1998}, wherein we let $\Psi_N(\theta) = \Psi_s(\theta)$. Note that this definition for $\Psi_N(\theta)$ differs from that provided in Section \ref{sec:intest}, which applies to the rest of this paper outside the current proof.
\end{proof}

\begin{proof}[Proof of Corollary \ref{crl:conspop}]
Let $\Psi(\theta)$ be the limit of $\Psi_N(\theta)$ and rearrange $\Psi_s(\theta)$ to give
\begin{equation*}
\Psi_s(\theta) = \Psi(\theta) + \frac{1}{N} \sum_{i=1}^N (w_i - 1) \psi(Y_i ; \theta) + (\Psi_N(\theta) - \Psi(\theta)).
\end{equation*}
The second and third terms are $o_P(1)$ by assumption, so $\Psi_s(\theta) \overset{P}{\to} \Psi(\theta)$.
\end{proof}

\begin{proof}[Proof of Theorem \ref{thm:normsup}]
Apply Theorem 3.3.1 of \citepref{vanderVaart1996}, noting that both $\Psi_s$ and $\Psi$ are defined on and take values in Euclidean space, so the paragraph preceding Theorem 3.3.1 applies.
\end{proof}

\begin{proof}[Proof of Corollary \ref{crl:normpop}]
Obtain \eqref{eqn:linpop} by subtracting $w$-weighted \eqref{eqn:linsup} by $\mathbf{1}$-weighted \eqref{eqn:linsup} as follows:
\begin{align*}
\sqrt{N}\dot{\Psi}_{\theta_0} (\hat{\theta}_s - \hat{\theta}_N) &= \sqrt{N} \dot{\Psi}_{\theta_0} (\hat{\theta}_s - \theta_0) - \sqrt{N} \dot{\Psi}_{\theta_0} (\hat{\theta}_N - \theta_0) \\
&= -\sqrt{N}(\Psi_s(\theta_0) - \Psi(\theta_0)) + o_P(1) + \sqrt{N}(\Psi_N(\theta_0) - \Psi(\theta_0)) - o_P(1) \\
&= - \sqrt{N}(\Psi_s(\theta_0) - \Psi_N(\theta_0)) + o_P(1).
\end{align*}
The continuous mapping theorem then gives \eqref{eqn:convdistprime}.
\end{proof}

\begin{proof}[Proof of Theorem \ref{thm:desvar}]
To obtain \eqref{eqn:designvar}, first note that,
\begin{align*}
\mathrm{Var} & \left( \mathbb{E} \left[ \sqrt{N}(\Psi_s(\theta_0) - \Psi_N(\theta_0)) \mid Y_{1:N} \right] \right)  \\
&= \mathrm{Var} \left( \sqrt{N}( \mathbb{E} \left[ \Psi_s(\theta_0) \mid Y_{1:N} \right] - \Psi_N(\theta_0)) \right) \\
&= \mathrm{Var} \left( \sqrt{N}(\Psi_N(\theta_0) - \Psi_N(\theta_0)) \right) \\
&= 0,
\end{align*}
then apply the law of total variance to $\mathrm{Var} \left( \sqrt{N}(\Psi_s(\theta_0) - \Psi_N(\theta_0)) \right)$, conditioning on $Y_{1:N}$. By replacing ``$\Psi_N(\theta_0)$'' with ``$\Psi(\theta_0)$'', a similar argument leads to \eqref{eqn:uncondvar}, except that $\mathrm{Var} \left( \mathbb{E} \left[ \sqrt{N}(\Psi_s(\theta_0) - \Psi(\theta_0)) \mid Y_{1:N} \right] \right) = N \mathrm{Var} \left( \Psi_N(\theta_0) \right)$.

To obtain \eqref{eqn:designvartheta} and \eqref{eqn:uncondvartheta}, first consider the remainders
\begin{align*}
r_s &= \sqrt{N} \dot{\Psi}_{\theta_0} (\hat{\theta}_s - \theta_0) + \sqrt{N} (\Psi_s(\theta_0) - \Psi(\theta_0)), \\
r_N &= \sqrt{N} \dot{\Psi}_{\theta_0} (\hat{\theta}_N - \theta_0) + \sqrt{N} (\Psi_N(\theta_0) - \Psi(\theta_0)).
\end{align*}
The remainder in \eqref{eqn:linpop} is then given by $r^{\prime}_N = r_s - r_N$; see the display in the proof of Corollary \ref{crl:normpop} above. By the triangle inequality, each element $r_{N i}$ satisfies $\mathbb{E}[(r^{\prime}_{N i})^2]^{1/2} \leq \mathbb{E}[(r_{s i})^2]^{1/2} + \mathbb{E}[r^2_{N i}]^{1/2}$, so $\mathbb{E}[(r^{\prime}_{N i})^2] \to 0$ by the assumed convergence of $\mathbb{E}[(r_{s i})^2]$ and $\mathbb{E}[r^2_{N i}]$ and the continuity of $x \mapsto x^2$ and $x \mapsto x^{1/2}$.

We also have
\begin{align*}
\mathrm{Var}(\mathbb{E}[r_{N i} \mid Y_{1:N}]) &\leq \mathbb{E}[\mathbb{E}[r_{N i} \mid Y_{1:N}]^2] \\
&\leq \mathbb{E}[ \mathbb{E}[r^2_{N i} \mid Y_{1:N}] ] \\
&= \mathbb{E}[r^2_{N i}] \\
&= o(1),
\end{align*}
where the second inequality follows from Jensen's inequality and the first equality from the law of total expectation. The Cauchy-Schwarz inequality then gives
\begin{equation*}
\mathrm{Cov}(\mathbb{E}[r_{N i} \mid Y_{1:N}], \mathbb{E}[r_{N j} \mid Y_{1:N}])^2 \leq \mathrm{Var}(\mathbb{E}[r_{N i} \mid Y_{1:N}]) \mathrm{Var}(\mathbb{E}[r_{N j} \mid Y_{1:N}]) = o(1),
\end{equation*}
so that
\begin{equation*}
\mathrm{Var}(\mathbb{E}[r_N \mid Y_{1:N}]) = o(1).
\end{equation*}
The above also applies to $r_s$ and $r^{\prime}_N$ in place of $r_N$.

In pursuit of \eqref{eqn:designvartheta}, it follows that
\begin{align*}
\mathrm{Var} & \left( \mathbb{E} \left[ \sqrt{N} \dot{\Psi}_{\theta_0} (\hat{\theta}_s - \hat{\theta}_N) \mid Y_{1:N} \right] \right) \\
&= \mathrm{Var} \left( \mathbb{E} \left[ - \sqrt{N}(\Psi_s(\theta_0) - \Psi_N(\theta_0)) + r^{\prime}_N \mid Y_{1:N} \right] \right) \\
&= \mathrm{Var} \left( - \sqrt{N} \left( \mathbb{E} \left[ \Psi_s(\theta_0) \mid Y_{1:N} \right] - \Psi_N(\theta_0) \right) + \mathbb{E}[r^{\prime}_N \mid Y_{1:N}] \right) \\
&= \mathrm{Var} \left( \mathbb{E}[r^{\prime}_N \mid Y_{1:N}] \right) \\
&= o(1).
\end{align*}
Now apply of the law of total variance to $\sqrt{N} \dot{\Psi}_{\theta_0} (\hat{\theta}_s - \hat{\theta}_N)$, premultiply by $\dot{\Psi}_{\theta_0}^{-1}$ and postmultiply by $(\dot{\Psi}_{\theta_0}^{-1})^T$, which gives \eqref{eqn:designvartheta} in view of the continuous mapping theorem.

To obtain \eqref{eqn:uncondvartheta}, first note that by the Cauchy-Schwarz inequality,
\begin{align*}
\mathrm{Cov} & \left( \sqrt{N}(\Psi_N(\theta_0)_i - \Psi(\theta_0)_i), \mathbb{E}[r_{s j} \mid Y_{1:N}] \right)^2 \\
&\leq \mathrm{Var} \left( \sqrt{N}(\Psi_N(\theta_0)_i - \Psi(\theta_0)_i) \right) \mathrm{Var} \left( \mathbb{E}[r_{s j} \mid Y_{1:N}] \right) \\
&= o(1),
\end{align*}
where convergence (and hence asymptotic boundedness) of the first variance follows after applying the asymptotic uniform integrability and convergence in distribution of its contents squared to Theorem 2.20 of \citepref{vanderVaart1998}. Thus
\begin{equation*}
\mathrm{Cov}\left( \sqrt{N}(\Psi_N(\theta_0) - \Psi(\theta_0)), \mathbb{E}[r_s \mid Y_{1:N}] \right) = o(1).
\end{equation*}
It follows that
\begin{align}
\mathrm{Var} & \left( \mathbb{E} \left[ \sqrt{N} \dot{\Psi}_{\theta_0} (\hat{\theta}_s - \theta_0) \mid Y_{1:N} \right] \right) \nonumber \\
=& \mathrm{Var} \left( \mathbb{E} \left[ - \sqrt{N} (\Psi_s(\theta_0) - \Psi(\theta_0)) + r_s \mid Y_{1:N} \right] \right) \nonumber \\
=& \mathrm{Var} \left( - \sqrt{N}(\mathbb{E}[\Psi_s(\theta_0) \mid Y_{1:N}] - \Psi(\theta_0)) + \mathbb{E}[r_s \mid Y_{1:N}] \right) \nonumber \\
=& \mathrm{Var} \left( - \sqrt{N}(\Psi_N(\theta_0) - \Psi(\theta_0)) + \mathbb{E}[r_s \mid Y_{1:N}] \right) \nonumber \\
=& \mathrm{Var} \left(  - \sqrt{N}(\Psi_N(\theta_0) - \Psi(\theta_0)) \right) + \mathrm{Var} \left( \mathbb{E}[r_s \mid Y_{1:N}] \right) \nonumber \\
&- \mathrm{Cov} \left( \sqrt{N}(\Psi_N(\theta_0) - \Psi(\theta_0)), \mathbb{E}[r_s \mid Y_{1:N}] \right) \nonumber \\
&- \mathrm{Cov} \left(\mathbb{E}[r_s \mid Y_{1:N}], \sqrt{N}(\Psi_N(\theta_0) - \Psi(\theta_0)) \right) \nonumber \\
=& \mathrm{Var} \left(  - \sqrt{N}(\Psi_N(\theta_0) - \Psi(\theta_0)) \right) + o(1). \label{eqn:expcondequncond}
\end{align}
Applying the above with $w_i = 1$ gives
\begin{equation*}
\mathrm{Var} \left( \sqrt{N} \dot{\Psi}_{\theta_0} (\hat{\theta}_N - \theta_0) \right) = \mathrm{Var} \left( - \sqrt{N}(\Psi_N(\theta_0) - \Psi(\theta_0)) \right) + o(1),
\end{equation*}
and by substituting the left-hand side of the above into the right-hand side of \eqref{eqn:expcondequncond}, we obtain
\begin{equation*}
\mathrm{Var} \left( \mathbb{E} \left[ \sqrt{N} \dot{\Psi}_{\theta_0} (\hat{\theta}_s - \theta_0) \mid Y_{1:N} \right] \right) = \mathrm{Var} \left(  \sqrt{N} \dot{\Psi}_{\theta_0} (\hat{\theta}_N - \theta_0) \right) + o(1).
\end{equation*}
Now apply the law of total variance to $\sqrt{N} \dot{\Psi}_{\theta_0} (\hat{\theta}_s - \theta_0)$, premultiply by $\dot{\Psi}_{\theta_0}^{-1}$ and postmultiply by $(\dot{\Psi}_{\theta_0}^{-1})^T$, which gives \eqref{eqn:uncondvartheta} in view of the continuous mapping theorem.
\end{proof}

\begin{proof}[Proof of Theorem \ref{thm:normweights}]
If the first equality holds, then
\begin{equation*}
\frac{1}{N} \sum_{i=1}^N w^{\prime}_i \psi(Y_i; \hat{\vartheta}_N) = \frac{\frac{1}{N} \sum_{i=1}^N w_i \psi(Y_i; \hat{\vartheta}_N)}{\frac{1}{N} \sum_{j=1}^N w_j} = \frac{o_P(r_N)}{1 + o_P(1)} = o_P(r_N),
\end{equation*}
and we obtain the second equality. If the second equality holds, then
\begin{equation*}
\frac{1}{N} \sum_{i=1}^N w_i \psi(Y_i; \hat{\vartheta}_N) = \left( \frac{1}{N} \sum_{j=1}^N w_j \right) \left( \frac{1}{N} \sum_{i=1}^N w^{\prime}_j \psi(Y_i ; \hat{\vartheta}_N) \right) = (1 + o_P(1)) o_P(r_N) = o_P(r_N),
\end{equation*}
and we obtain the first equality.
\end{proof}

\if0\wholepaper {
	\bibliographypref{../library}
} \fi